\documentclass[12pt]{article}
\usepackage{amsmath}
\usepackage{graphicx,psfrag,epsf}
\usepackage{enumerate}
\usepackage{natbib}
\usepackage{url} % not crucial - just used below for the URL 

\newcommand{\blind}{0}

  \usepackage{setspace}

\begin{document}
	
%\bibliographystyle{natbib}
	
%%%%%%%%%%%%%%%%%%%%%%%%%%%%%%%%%%%%%%%%%%%%%%%%%%%%%%%%%%%%%%%%%%%%%%%%%%%%%%

\def\spacingset#1{\renewcommand{\baselinestretch}%
	{#1}\small\normalsize} \spacingset{1}

\if0\blind
{
	\title{ \bf Poisson Kernel-Based Clustering on the Sphere: Convergence Properties, Identifiability, and a Method of Sampling}
	\author{Mojgan Golzy \vspace{.2cm} and Marianthi Markatou \thanks{The second author gratefully acknowledges financial support provided by the Department of Biostatistics, University at Buffalo in the form of a start-up package that funded the work of the first author. } \\
	  Department of Biostatistics, University at Buffalo, Buffalo, NY\\
		 	}
	\maketitle
} \fi

\if1\blind
{
	\bigskip
	\bigskip
	\bigskip
	\begin{center}
		{ \bf Poisson Kernel-Based Clustering on the Sphere: Convergence Properties, Identifiability, and a Method of Sampling}
	\end{center}
	\medskip
} \fi

\bigskip
\begin{abstract}
Many applications of interest involve data that can be analyzed as unit vectors on a {\it d}-dimensional sphere. Specific examples include text mining, in particular clustering of documents, biology, astronomy and medicine among others. Previous work has proposed a clustering method using mixtures of Poisson kernel-based distributions (PKBD) on the sphere. 
We prove identifiability of mixtures of the aforementioned model, convergence of the associated EM-type algorithm and study its operational characteristics. Furthermore, we propose an empirical densities distance plot for estimating the number of clusters in a PKBD model. Finally, we propose a method to simulate data from Poisson kernel-based densities and exemplify our methods via application on real data sets and simulation experiments. \\

\end{abstract}

\noindent%

\vfill
	
\newpage
\spacingset{1.45} % DON'T change the spacing!

\section{Introduction}
\label{sec:intro}
	
Directional data arise naturally in many scientific fields where observations are recorded as directions or angles relative to a fixed orientation system. Directions may be regarded as points on the surface of a hypersphere, thus the observed directions are angular measurements. Directional data are often met in astronomy, where the origin of comets is investigated or in biology, where clustering of gene expression measurements that are standardized to have  mean zero and variance 1 across arrays is of interest. \citet{Jamma} discuss a problem in medicine where the angle of knee flexion was measured to assess the recovery of orthopaedic patients. Furthermore, \cite{peel}
discuss the analysis of directional data in an application in the mining industry, where a mine tunnel is modeled. 

Conventional methods suitable for the analysis of linear data cannot be applied for directional data due to its circular nature. The statistical methods that are used to handle such data are given in several references such as \citet{Watson, Fisher, Mardia,Lee}.  
Clustering methods for directional data have been developed in the literature. 
Some commonly used non-parametric approaches are {\it K}-means clustering \citep{Ramler,Maitra}, spherical {\it K}-means \citep{Dhillon}, and online spherical {\it K}-means \citep{Zhong}. 
Furthermore, some of the clustering methods proposed in the literature are appropriate for small and medium dimensional data sets, while high dimensional data are considered in \citet{Dryden,Ban03a, Ban05, Zhong03,Zhong05}, with applications to brain shape modeling, text data represented by large sparse vectors, and genomic data.

Probability models have been proposed for quite sometime as a basis for cluster analysis.  In this approach the data are viewed as generated from  a mixture of probability distributions, each representing a different cluster. Clustering algorithms based on probability models allow uncertainty in cluster membership, and direct control over the variability allowed within each cluster. Probabilistic approaches are also called generative approaches and a list of references on these approaches in the context of clustering text can be found in \citet{Zhong03,Zhong05} and  \citet{Blei03}.
\citet{Ban05} considered a finite mixture of von Mises-Fisher (vMF) distributions to cluster text and genomic data.  The spherical k-means algorithm, has been shown to be a special case of a generative model based on a mixture of vMF distributions with equal priors for the components and equal concentration parameters \citep{Ban02,Ban03a}.  A comparative study of some generative models based on the multivariate Bernoulli, multinomial distributions, and the generative model based on a mixture of vMF distributions is presented in \citet{Zhong03}.

\cite{Golzy}, presented a clustering algorithm based on mixtures of Poisson kernel-based distributions (PKBD). Poisson kernels on the sphere \citep{Lindsay02} have important mathematical and physical interpretation.  A clustering algorithm was devised and estimates of the parameters of the Poisson kernel-based algorithm were obtained in an Expectation-Maximization (EM) setting.  Experimental and simulation results indicated that the method performs at least equivalently to the mixture of vMF distributions, which is considered to be the state of the art, and outperforms the aforementioned algorithm in certain data structures, when performance is measured by macro-precision and macro-recall. 

In this paper, we present a detailed study of our clustering algorithm, investigate its properties and illustrate its performance.  Specifically, our contributions are as follows. First, we study the connection between PKBD and other spherical distributions. Section 3.2 presents the results of the aforementioned study. Section 4 of the paper establishes the identifiability of a mixture model of Poisson kernel-based densities, a new contribution in establishing validity of our PKBD algorithm. Section 5 establishes the convergence of our proposed algorithm, while section 6 discusses a method of sampling from a PKBD family. Practical issues of implementation of our algorithm such as study of the role of initialization on the performance of the algorithm, stopping rules and a method for estimating the number of clusters when data are generated from a mixture of PKBD distributions 
are discussed in section 7.  Section 8 presents experimental results that illustrate the performance of our algorithm, while section 9 offers discussion and conclusions. The online supplemental material associated with the paper contains detailed proofs of our theoretical results and additional simulations, illustrating further the performance of the algorithm. The code and data sets are also provided in the online supplement.

\section{Literature Review}
In this section, we briefly review the clustering literature for directional data. We, very briefly, refer to algorithms that are distance or similarity based (non-generative algorithms) while our focus is on probabilistic (or generative) algorithms. We begin with a brief description of non-generative algorithms for directional data.

{\it K}-means clustering \citep{Duda} is one of the most popular methods for clustering. Given a set ${\cal X}$ of $N$ observations, where each observation is a $d$-dimensional real vector, {\it K}-means clustering  partitions the $N$ observations into $M (\leq N)$ sets ${\cal X}_1, \cdots, {\cal X}_M$ by minimizing the within-cluster sum of squares.  Spherical {\it K}-means  \citep{Dhillon}, uses cosine similarity instead of Euclidean distance, that measures the cosine of the angle formed by two vectors. Spherical {\it K}-means algorithm is preferred to standard {\it K}-means for clustering of  document vectors or any type of high-dimensional data on the unit sphere, and it is sensitive to initialization and outliers. 

\citet{Maitra} propose a {\it K}-means directions algorithm for fast clustering of data on the sphere. They modified the core elements of \citet{Hartigan} efficient {\it K}-means implementation for application to spherical data.  Their algorithm incorporates the additional constraint of orthogonality to the unit vector, and thus extends to the situation of clustering using the correlation metric.

\subsection{Parametric Mixture Model Approach for Clustering} 

The parametric mixture model assumes each cluster is generated by its own density function that is unknown. The overall data is modeled as a mixture of individual cluster density functions. In practice, the unknown densities may not be from the same family of distributions. In this section, we consider mixture models in which the densities are from the same family of distributions.  The probability density function of a mixture with $M$ components on the hypersphere $ S^{d-1}$, the unit sphere,  is given by
\begin{equation} \label{eq:a1}
	f({\bf x}|\Theta)=\sum_{j=1}^M \alpha_j  f_j({\bf x}|\mbox{\boldmath$\theta$}_j), 
\end{equation}
where $M$ is the number of clusters, $\alpha_j$'s are the mixture proportions that are non-negative and sum to one and $\Theta=(\alpha_1, \cdots, \alpha_M, \mbox{\boldmath$\theta$}_1, \cdots, \mbox{\boldmath$\theta$}_M)$.

\citet{Ban05} discuss clustering based on mixtures of von Mises-Fisher (vMF) distributions on a hypersphere.  Given $\mbox{\boldmath$\mu$}\in S^{d-1},$ and $\kappa\geq 0$, the vMF probability distribution function is defined by
$		f({\bf x}|\mbox{\boldmath$\mu$}, \kappa )= c_d(\kappa) e^{\kappa \mbox{\boldmath$\mu$} {\bf \cdot} {\bf x}},
$
where $\mbox{\boldmath$\mu$}$ is a vector orienting the center of the distribution, $\kappa$  is a parameter to control the concentration of the distribution around the vector $\mbox{\boldmath$\mu$} $ and ${\bf y} {\bf \cdot}{\bf x}$ denote the dot product of the vectors. The normalizing constant $c_d(\kappa)$ is given by
$ c_d(\kappa)=\frac{\kappa^{d/2-1}}{(2\pi)^{d/2} I_{d/2-1}(\kappa)},$
where $I_r(.)$ represents the modified Bessel function of the first kind of order $r$. The vMF distribution is unimodal and symmetric about  $\mbox{\boldmath$\mu$}$. 

\citet{Ban05} performed Expectation Maximization (EM) \citep{Dempster, Bilmes} for a finite vMF mixture model to cluster text and genomic data. 	The numerical estimation of the concentration parameter  involves functional inversion of the ratios of Bessel functions.  Thus,  it is not possible to directly estimate the $\kappa$ values in high dimensional data and an asymptotic approximation of $\kappa$ is used for estimating $\kappa$.  
The package movMF in R software can be used for fitting a mixture of vMF distribution (Hornik and Gr$\ddot{\rm u}$n 2014).
	
Mixtures of Watson distributions are discussed in \citet{Bijral} and \citet{Sra}. Given $\mbox{\boldmath$\mu$}\in S^{d-1}$ and $\kappa$, the probability function of a Watson distribution  is defined by
$
		f({\bf x}| \mbox{\boldmath$\mu$}, \kappa)=M(1/2, d/2, \kappa)^{-1} 
		e^{\kappa (\mbox{\boldmath$\mu$} {\bf \cdot} {\bf x})^2},
$
where $M(1/2, d/2, \kappa)$ is the confluent hyper-geometric function also known as Kummer function.  The advantage of using the class of Watson distributions in the mixture model is that it shows superior performance, when the measure of performance is the mutual information between cluster assignment and preexisting labels, for noisy, thinly spread clusters over the vMF distributions \citep{Bijral}.  The disadvantage is that in high-dimensions, maximum likelihood equations pose severe numerical challenges. Similar to vMF, it is not possible to directly estimate the $\kappa$ values, since the numerical estimation of $\kappa$ involves a ratio of Kummer functions, and hence an asymptotic approximation for estimating  $\kappa$ is used.

\citet{Dortet} have presented model based clustering of data on the sphere by using  inverse stereographic projections of multivariate normal distributions.  Recall that, given a direction $\mbox{\boldmath$\mu$}$ on the sphere $S^{d-1}$, the corresponding stereographic projection of a point  ${\bf x}$ that belongs to $S^{d-1}$ lies at the intersection of a line joining the "antipole" $-\mbox{\boldmath$\mu$}$ and ${\bf x}$, with a given plane perpendicular to $\mbox{\boldmath$\mu$}$. 
Let ${\cal L}_{\mu, \Sigma}$ denote the distribution on the sphere $S^{d-1}$, which corresponds to the image via an inverse stereographic projection of a multivariate normal distribution $N_{d-1}({\bf 0}, \Sigma)$ that is defined on the plane of dimension $d-1$ perpendicular to $\mbox{\boldmath$\mu$}$. 
The density function of ${\cal L}_{\mu, \Sigma}$ is given by
	\begin{equation}		f_{\mu, \Sigma}({\bf x})=\frac{1}{(2\pi)^{(d-1)/2}}
|\Sigma|^{-1/2} \exp \{     -1/2 P(R_{\mu^{-1}}({\bf x}))^T \Sigma^{-1} P(R_{\mu^{-1}}({\bf x}))   \}
 \frac{1}{(1+\mbox{\boldmath$\mu$} {\bf \cdot} {\bf x})^{d-1}}, 	\end{equation}
where $P(.)$ is the stereographic projection map and $R_{\mu}(.)$ is the rotation in ${\rm I\!R}^d$ such that $   R_{\mu}(e_1)=\mbox{\boldmath$\mu$}$, where
$\{ e_1, \cdots, e_d\}$ is the canonical basis of the  ${\rm I\!R}^d$. Given $\mbox{\boldmath$\mu$}$, 
$   \hat \Sigma_\mu =1/n \sum_{i=1}^n  P(R_\mu^{-1}({ {\bf x}_i})) 
P(R_\mu^{-1}({ {\bf x}_i}))^T,$
and $\mbox{\boldmath$\mu$}_{MLE}$ maximizes the expression given by\\
${\rm Expr}(\mbox{\boldmath$\mu$})=-1/2n \log(|\hat \Sigma_\mu |)- (p-1)\sum_{i=1}^n \log(1+\mbox{\boldmath$\mu$} {\bf \cdot} {\bf x_i}).$

The advantage of using the class of inverse stereographic projection of the multivariate normal distribution in the mixture model is that it
allows clustering with various shapes and orientations. The projected multivariate normal is applied to a real data set of standardized gene expression profiling. The disadvantage is that, there is no closed expression for $\mbox{\boldmath$\mu$}_{MLE}$. In practice, it is obtained via a heuristic search algorithm.

\section{Clustering Based on Mixtures of Poisson Kernel-Based Distributions}
\label{sec:PKBD}

We propose a parametric mixture model approach to clustering directional data based on Poisson kernel-based distributions  on the unit sphere. Clustering on the basis of Poisson kernel-based densities avoids the use of approximations, obtains closed form solutions and provides robust clustering results.

\subsection{Poisson Kernel-Based Distributions (PKBD)}

We use Poisson kernel as a density function on the sphere.  To provide perspective  we note here that the simplest PKBD provided by the univariate Poisson kernel, is a circular distribution that is also known as the wrapped Cauchy distribution.  This distribution can be constructed by "wrapping" the univariate Cauchy distribution around the  circumference of the circle of unit radius. It was studied first by  \citet{Levy} and \citet{Wintner}.

Let ${\rm I\!B}^d$   be the  open unit ball in ${\rm I\!R}^d$ (i.e; ${\rm I\!B}^d(0,1)$) and $ S^{d-1}$ be the unit sphere, The $d$-dimensional {\it Poisson kernel for the unit ball} is defined for 
$({\bf x}, \mbox{\boldmath$\zeta$})\in {\rm I\!B}^d\times S^{d-1} $ by
\begin{equation} P_d({\bf x}, \mbox{\boldmath$\zeta$})=\frac{1-\big \|{\bf x}\big \|^2}{\omega_d \big \|{\bf x}-  \mbox{\boldmath$\zeta$}\big \|^d} ,\end{equation}
where $\omega_d=2 \pi^{d/2}\{\Gamma(d/2)\}^{-1}$ is the surface area of the unit sphere in ${\rm I\!R}^d$. The family of Poisson kernels is the set $\{K_\rho({\bf x},{\bf y}):  0<\rho <1 \} $, where $ K_\rho({\bf x},{\bf y})$ defined on  $ S^{d-1} \times S^{d-1}$ by  
\begin{equation} \label{K} K_\rho({\bf x},{\bf y})= P_d(\rho{\bf x}, {\bf y}).\end{equation}
Let $\sigma$ be the uniform measure on $S^{d-1}$ (so that $\sigma(S^{d-1})=\omega_d$), then
$\int_{S^{d-1}} K_\rho({\bf x}, \mbox{\boldmath$\zeta$})d \sigma(\mbox{\boldmath$\zeta$})=1,$ and so $K_\rho({\bf x},{\bf y})$ is a density with respect to uniform measure \citep{Axler,Lindsay02,Dai}.

We discuss clustering based on mixtures of Poisson kernel-based distributions (mix-PKBD) on a hypersphere.  Given $\mbox{\boldmath$\mu$}\in S^{d-1},$ and $0<\rho<1$, the probability distribution function of a d-variate Poisson kernel-based density is defined by
\begin{equation} \label{eq:Poisson} 
f({\bf x}|\rho, \mbox{\boldmath$\mu$})= \frac{1-\rho^2}{\omega_d \big \|{\bf x}-\rho  \mbox{\boldmath$\mu$}\big \|^d},
\end{equation}
where $\mbox{\boldmath$\mu$}$ is a vector orienting the center of the distribution, and $\rho$  is a parameter to control the concentration of the distribution around the vector $\mbox{\boldmath$\mu$} $. That is, the parameter $\rho$ is related to the variance of the distribution. PKBDs are unimodal and symmetric around $ \mbox{\boldmath$\mu$} $. Figure 1 shows the shape of Poisson kernel-based densities for various values of the parameter $\rho$.  For additional pictorial representation of the PKBDs for various values of $\rho$ see Figure A15 of the supplemental material. We note that 
\begin{equation} \label{eq:limits}
\frac{1-\rho}{\omega_d (1+\rho)^{d-1}}< f({\bf x}|\rho, \mbox{\boldmath$\mu$}) <\frac{1+\rho}{\omega_d (1-\rho)^{d-1}}.	\end{equation}
Therefore, if $\rho \rightarrow 0 $ then  $f({\bf x}|\rho, \mbox{\boldmath$\mu$}) \rightarrow 1/\omega_d $ which is the uniform density on  $S^{d-1}$ and if $\rho \rightarrow 1$,  $f({\bf x}|\rho, \mbox{\boldmath$\mu$})$ converges to a point density.  
 
 \begin{figure}[h]
 	\begin{center}
 		\includegraphics[width=5.5in]{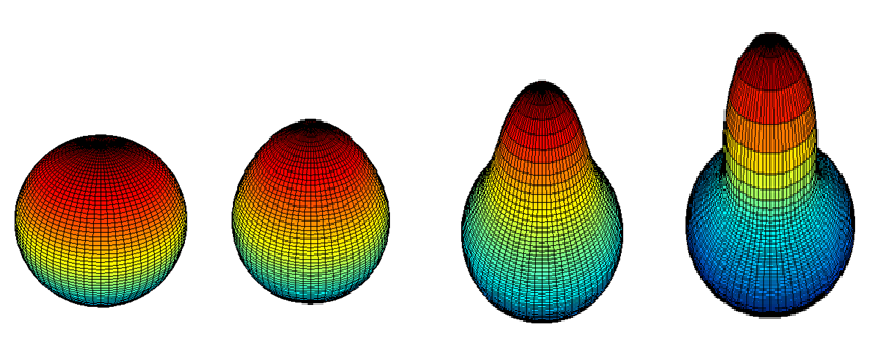}
 	\end{center}
 	\caption{Scaled 3-variant Poisson kernel-based density with $\mbox{\boldmath$\mu$}=(0,0,1)$ and various $\rho$ values.}
 \end{figure}

\subsection{Connections with Other Spherical Distributions}
	
In general, given a distribution on the line, \citet{Mardia} note that we can wrap it around 	the circumference of the circle of unit radius. 
 If $X$ has distribution $F$ then the wrapped distribution $F_w$ of $\theta$ is given by
	\begin{equation} 
		F_w(\theta)= \sum_{k=-\infty}^{k=\infty}\{F(\theta+2 \pi k)-F(2 \pi k)\}  {\rm ~ for ~} 0\leq \theta \leq 2\pi,\end{equation} 
where $\theta= X \mod 2\pi $.  	In particular if $\theta$ has density $f$ then $f_w(\theta)=\sum_{k=-\infty}^{k=\infty} f(\theta+2 \pi k)$ \citep{Mardia}.\\
	
\citet{Mardia} note that both  wrapped normal and wrapped Cauchy (that is PKBD for d=2) can be use as an approximation of vMF distributions. 	Figure 2, gives the plots of the two distributions, PKBD (solid line) and vMF (dashed line), with the same mean values and various $\kappa$ and $\rho$ values. The corresponding $\rho$ values are chosen in a way that both distributions have the same maximum. The variable $t$ is the angle between ${\bf x}$ and $\mbox{\boldmath$\mu$}$, measured in radian (from -3.14 to 3.14 radians). We note that the PKBD has heavier tails than the vMF distribution.  We will illustrate this fact for dimension 4 in the supplemental material (Figure A16).  PKBD also has heavier tails than the  Elliptically Symmetric Angular Gaussian (ESAG) \citep{Paine} which is a subfamily of Angular Central Gaussian Distribution (ACGD). An illustration is given in the supplemental material (Figure A17). Furthermore, notice that as $\kappa$ increases the value of $\rho$ also increases.
 \begin{figure}[h!]
 	\centering
 	\includegraphics[width=1\textwidth]{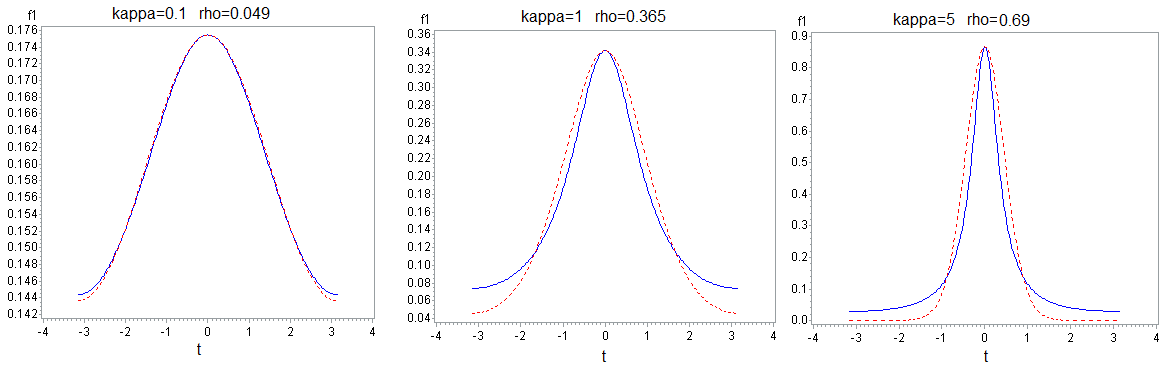}
 	\caption {Comparison of the Poisson kernel-based (solid) and von Mises Fisher (dashed) distributions for d=2 with the same maximum values, $t$ is the angle between ${\bf x}$ and $\mbox{\boldmath$\mu$}$ measured in radian. }
 \end{figure}
 
The two dimensional PKBD is also related to projected normal distribution. 
	Let ${\bf y}=(y_1, \cdots,y_d)\sim N_d(\mbox{\boldmath$\mu$}, \Sigma)$ with $P({\bf y}=0)=0$ then ${\bf u}={\bf y}/|{\bf y}|$ is a random variable in $S^{d-1}$. The random variable  ${\bf u}$ has a projected normal distribution denoted by ${\bf u}\sim PN_d(\mbox{\boldmath$\mu$}, \Sigma)$.
	In the special case where $\mbox{\boldmath$\mu$}=0$,  the density of ${\bf u}$ is given by
\begin{equation}f({\bf u}| \mbox{\boldmath$\mu$}=0, \Sigma) = \frac{1}{\omega_d |\Sigma|^{1/2}({\bf u}^t \Sigma^{-1}{\bf u})^{d/2}}.\end{equation}
	This is the {\it angular central Gaussian distribution} \citep{Mardia, Paine}.
		
\citet{Mardia} show that if $\theta $ is a random vector that follows a 2-dimensional 
 $ PN_2( \mbox{\boldmath$\mu$}=0, \Sigma)$, where $\Sigma=(\sigma_{ij})_{i,j=1,2} $,
	then $2\theta $ follows a PKBD with parameters given as
	\[\rho=\left\lbrace \frac{tr(\Sigma)-2 |\Sigma|^{1/2}}{tr(\Sigma)+2 |\Sigma|^{1/2}}\right\rbrace^{1/2}, \  \mu= e^{i\alpha}, \ \alpha=\tan^{-1} \left\lbrace \frac{2 \sigma_{12}}{\sigma_{11}-\sigma_{22}}\right\rbrace. \]

This connection of the two-dimensional projected normal family with the PKBD family cannot be extended beyond $d=2$. Below, we provide a specific example of a $d$-dimensional projected normal family for which  $R^d_\theta {\bf u}$, ${\bf u} \in S^{d-1}$ does not follow a PKBD for $d>2$,  $R^d$ is the rotation matrix through $\theta$.\\

	 \noindent
	{\it Proposition 3.1.}  Let ${\bf u} $ be a random vector in $S^{d-1}$ with mean zero projected normal density $ PN_d( \mbox{\boldmath$\mu$}=0, \Sigma)$, with
	\[ 
	\sigma_{ij} = 
	\left\{
	\begin{array}{ll}
	0
	&\mbox{if }  i\neq j\\
	1 & \mbox{if }i=j=2, \cdots, d. \\
	\sigma^2\neq1 & \mbox{if }i=j=1. \\
	\end{array}\right.
	\]
	Then, $R^d_\theta {\bf u}$ has Poisson kernel-based density if and only if $d=2$, where $R^d_\theta$ is the rotation matrix through the angle $\theta$. 
	
	{\it Proof.} Given in the online supplementary materials.

\subsection{Estimation of the Parameters of the Mixtures of PKBD}
Let $ {\cal X}$ be a set of sample unit vectors drawn independently from mixtures of Poisson kernel-based distributions. Our model is a mixture of $M$ Poisson kernel-based densities with parameters $(\alpha_j, \rho_j, \mbox{\boldmath$\mu$}_j)$, where $\alpha_j$ corresponds to the  weights of the mixture components \& $\rho_j, \mbox{\boldmath$\mu$}_j,$ $j=1, \cdots, M$  are individual density based parameters. Thus, the parameter space  $\Theta=(\alpha_1, \cdots, \alpha_M, \rho_1, \cdots, \rho_M, \mbox{\boldmath$\mu$}_1, \cdots, \mbox{\boldmath$\mu$}_M)$, where $M$ is the number of clusters,  $\alpha_j \geq 0, j=1, \cdots, M$ and $\sum_{j=1}^M \alpha_j =1$.

The expectation of the complete likelihood is given as
 {\small \begin{equation} \label{eq:e15}
 	\sum_{j=1}^M\sum_{i=1}^N ln(\alpha_j)
 	p(j|{\bf x_i}, \Theta)+\sum_{j=1}^M\sum_{i=1}^N  ln(f_{j}({\bf x_i}|\rho_j, \mbox{\boldmath$\mu$}_j)p(j|{\bf x_i}, \Theta),
 	\end{equation}}
where  $p(j|{\bf x_i}, \Theta)$ is the posterior probability that ${\bf x_i}$ belongs to the $j^{th}$ component. 
 
The expression in (\ref{eq:e15}) contains two unrelated terms that can be separately maximized. From the maximization of the first term  in (\ref{eq:e15}) under the constraint $\sum_{j=1}^M \alpha_j=1$, given $\Theta^{(t-1)}$ we obtain
\begin{equation} \alpha_j^{(t)}=1/N \sum_{i=1}^N p(j|{\bf x_i},\Theta^{(t-1)}),	\end{equation}
where \begin{equation} p(j|{\bf x_i}, \Theta)=\frac{ \alpha_j  f_j({\bf x_i}|\rho_j, \mbox{\boldmath$\mu$}_j)}{\sum_{l=1}^M \alpha_l  f_l({\bf x_i}|\rho_l, \mbox{\boldmath$\mu$}_l) }.	\end{equation}
For details on EM algorithm we refer to \cite{Dempster,Bilmes}.
The  Lagrangian for the second term of (\ref{eq:e15})  is given by
{\small 	\begin{equation}
	\sum_{j=1}^M\sum_{i=1}^N   \{ ln(1-\rho_j^2) - ln(\omega_d)- d \  ln \big \|{\bf {\bf x_i}}-\rho_j \mbox{\boldmath$\mu_j$} \big \|\} \times
	p(j|{\bf x_i}, \Theta^{(t-1)})
	+\sum_{j=1}^M \lambda_j (1-\big \|\mbox{\boldmath$\mu$}_j\big \|^2).
	\end{equation}}
To estimate the parameters, we maximize the above expression, subject to $0<\rho<1$ for each $j$.

	 \noindent
	 {\it Proposition 3.2.}  The parameters ${\mbox{\boldmath$\mu_j$}}$, $\rho_j$ and $\alpha_j$, for $j=1, \cdots, M$, can be estimated using the iterative re-weighted algorithm given in Table 1.
	 
	 {\it Proof.} Given in the online supplementary materials.	
	
\begin{table}
	\caption{Algorithm for computing relevant estimates in a mixture of Poisson kernel-based density model.}
\noindent\fbox{%
	\parbox{\textwidth}{\small
		\begin{spacing}{0.8}
				\begin{itemize}
					\item {\bf Input:} Set ${\cal X}$ of data points on $S^{d-1}$, and  $M$ number of clusters.
					
					\item {\bf Output:} Clustering of ${\cal X}$ over a mixture of $M$ Poisson kernel-based  distributions.
					
					Initialize  $\alpha_k, \rho_k, \mbox{\boldmath$\mu$}_k$, for $ k= 1,\cdots , M$\\
					repeat
			\{E step\}
			\begin{itemize}
				\item for $k = 1$ to $M$ do
				\begin{itemize}
					\item for $i = 1$ to $n$ do			
					\hspace{2.5cm}\[h_k(x_i|\Theta_k) \leftarrow \frac{1-\rho_k^2}{\{ 1+\rho_k^2 -2 \rho_k {\bf x}_i.\mu_k\}^{d/2}}, \hspace{6cm}\]
					\item	end for
					\[p(k|{\bf x}_i,\Theta_k) \leftarrow \frac{\alpha_k h_k({\bf x}_i|\Theta_k)}{\sum_{l=1}^M \alpha_l h_l({\bf x}_i|\Theta_l)}, \hspace{8cm}\]
					\item for $i = 1$ to $n$ do
					\hspace{2.5cm} \[w_{ik} \leftarrow \frac{p(k|{\bf x}_i,\Theta_k)}{ 1+\rho_k^2 -2 \rho_k {\bf x}_i.\mu_k}, \hspace{7cm}\]
					\item end for
				\end{itemize}
				\item end for
			\end{itemize}
				\{M step\}
				\begin{itemize}
					\item for $k=1$ to $M$ do
					\[\alpha_k   \leftarrow 1/n \sum_{i=1}^n p(k|{\bf x}_i,\Theta_k), \hspace{8.4cm}\]
					\[\mbox{\boldmath$\mu$}_k  \leftarrow\frac{\sum_{i=1}^n w_{ik}{\bf x}_i}{\big \|\sum_{i=1}^n w_{ik}{\bf x}_i\big \|}, \hspace{9cm}\]
					\[\rho_k \leftarrow \rho_k- \frac{ g_k(\rho_k)}{ g_k'(\rho_k)},\hspace{9.2cm}\]
					\item end for
				\end{itemize}
				\item until converge
			\end{itemize}
			\end{spacing}
			}
}
\end{table}

\section{Identifiability of  Poisson Kernel-Based Mixtures of Distributions}
\label{sec:Iden}

Two kinds of identification problems are met when one works with mixture models; first we can always swap the labels of any two components with no effect on anything observable at all. Secondly, a more fundamental lack of identifiability happens when mixing  of two distributions from a parametric family just gives us a third distribution from the same family. 

\noindent{\it Definition 4.1.} \citep{Lindsay1995, Holzmann} Finite mixtures are said to be identifiable if distinct mixing distributions with finite support correspond to distinct mixtures. That is, finite mixtures from the family
$\{f(x,\theta_i):  \theta_i \in \Theta \}, $ are identifiable if 
\begin{equation} \sum_{j=1}^K \alpha_j f(x,\theta_j)= \sum_{j=1}^K \alpha'_j f (x,\theta_j'),\end{equation}
where $K$ is a positive integer, $\sum_{j=1}^K \alpha_j= \sum_{j=1}^K \alpha'_j=1$ and $ \alpha_j, \alpha'_j>0 $ for $j =1,\cdots ,K$, implies that there exists a permutation $\sigma$  such that $(\alpha_j', \theta_j' )= (\alpha_{\sigma(j)}, \theta_{\sigma(j)} )$ for all j.

Finite mixtures are identifiable if the family $\{f(x,\theta_i):  \theta_i \in \Theta \}, $ is linearly independent \citep{Yako}. That is, $\sum_{j=1}^K \alpha_j f(x,\theta_i)=0, \forall x$ implies $\alpha_j=0, $ $\forall j=1, \cdots, K$.

To prove the identifiability of a mixture of a Poisson kernel-based distributions we use the following representation of the Poisson kernel 
\begin{equation}\label{eq:14}	
K_\rho ({\bf x},\mbox{\boldmath$\mu$})=\frac{1}{\omega_d}\sum_{n=0}^\infty \rho^n Z_n({\bf x},\mbox{\boldmath$\mu$}), \end{equation}
where  $Z_n({\bf x},\mbox{\boldmath$\mu$})$ is a zonal harmonic \cite{Axler, Dai}. We then prove the following.

\noindent
{\it Lemma 4.2.} If  $ \sum_{j=1}^K \alpha_j K_{\rho_j}(x, \mu_j)=0 $, for all $x$, 
then $ \sum_{j=1}^K \alpha_j \rho_j^n Z_n(x,\mu_j) =0$, for each $x$ and $n$, where $Z_n(.,\mu_j)$ is the zonal harmonic of degree $n$ with pole $\mu_j$.

{\it Proof.} Given in the online supplementary materials.

\noindent 
{\it Proposition 4.3.}  Finite mixtures of the family $\{K_{\rho_j}(x,\mu_i): 0<\rho_j<1, \mu_j \in S^{d-1} \} $ of Poisson kernel-based distributions, are linearly independent. 

{\it Proof.} Given in the online supplementary materials.

\section{Convergence of the Algorithm}
\label{sec:conv}

To prove the convergence of the algorithm, we use a modification of the method used in \citet{Xu} and show that after each iteration the log-likelihood function increases. Since it is bounded, it is guaranteed to converge to a local maximum. 

\noindent {\it Theorem 5.1.} Let $\hat \Theta$ be the estimate obtained via the iterative EM algorithm given in Table 1. We use notations $ {\cal A}:=(\alpha_1, \cdots, \alpha_M)$ , 
${\cal M}:=(\mbox{\boldmath$\mu_1$}^T, \cdots , \mbox{\boldmath$\mu_M$}^T) $, and 
${\cal R}:=(\rho_1, \cdots, \rho_M)$.  At each iteration, we have:
\begin{itemize}
	\item [1.] \hspace{1cm} ${\cal A}^{(t)}- {\cal A}^{(t-1)}= {\cal P}_{\cal A}^{(t-1)}\frac {\partial l}{\partial {\cal A}}|_{{\cal A}={\cal A}^{(t-1)}},$
	
	where ${\cal P}_{\cal A}^{(t-1)}= 1/n \{{\rm diag} (\alpha_1^{(t-1)}, \cdots, \alpha_M^{(t-1)})- {\cal A}^{(t-1)}({\cal A}^{(t-1)})^T\}.$
	
	\item [2.]  \hspace{1cm} ${\cal M}^{(t)}- {\cal M}^{(t-1)}= {\cal P}_{\cal M}^{(t-1)}\frac {\partial l}{\partial {\cal M} }|_{{\cal M}= {\cal M}^{(t-1)}},$
	
	where ${\cal P}_{\cal M}^{(t-1)}= {\rm diag} (a_1^{(t-1)}I_d, \cdots,a_M^{(t-1)}I_d) $ and $a_k^{(t-1)}=\{  d\rho^{(t-1)}_k \big \|\sum_{i=1}^n w_{ik}^{(t-1)} {\bf x_i}\big \| \}^{-1}.$
	
	\item [3.]  \hspace{1cm} ${\cal R}^{(t)}- {\cal R}^{(t-1)}= {\cal P}_{\cal R}^{(t-1)}\frac {\partial l}{\partial {\cal R}}|_{{\cal R}={\cal R}^{(t-1)}},$
	
	where ${\cal P}_{\cal R}^{(t-1)}= {\rm diag}(\{- g_1'(\rho^{(t-1)}_1)\}^{-1},\cdots \{- g_M'(\rho^{(t-1)}_M)\}^{-1} ). $
\end{itemize}

{\it Proof.} Given in the online supplementary materials.

\noindent{\it Theorem 5.2.}  At each iteration of the EM algorithm, the direction of  $\Theta^{(t)}-\Theta^{(t-1)}$ has a positive projection on the gradient of the log-likelihood $l$.

{\it Proof.} Given in the online supplementary materials.

Therefore, the likelihood is guaranteed not to decrease after each iteration. Since $f({\bf x}|\Theta)$ is bounded (by \ref{eq:limits}), the log-likelihood function $l$ is bounded, and so, it is guaranteed to converge to a local maximum. 

\section{A Method of Sampling from a PKBD}
\label{sec:Gener}

To generate random samples from a two-dimensional PKBD we use the inverse sampling technique.
 Note that when d=2 the cumulative distribution function is 
\begin{equation} CDF(x)=\frac{1}{2\pi} \int_0^x \frac{(1-r^2)d\theta}{1+r^2-2 r \cos(\theta)}=\frac{1}{2\pi} {\rm arctg} \frac{(1+r) {\rm tg}(x/2)}{1-r}. \end{equation}

Finding an explicit formula for the inverse of the cumulative function of the PKBD for higher dimensions is not possible, and so inverse transform is not applicable.
For higher dimensions, we use the acceptance-rejection method for generating random variables from this distribution.

The basic idea is to find an alternative probability distribution $G(x)$, with density function $g(x)$,
for which we already have an efficient algorithm to generate data from, but also such that the function $g(x)$ is “close” to $f(x)$.
In particular, we assume that the ratio $f(x)/g(x)$ is bounded by a constant $M > 0$ (that is   $f(x) \leq Mg(x)$); we would want $M$ as close to 1 as possible. 

To generate a random variable $X$ from $F$, we first generate $Y$ from  $G$. Then generate $U \sim U(0,1)$ independent of $Y$. If $U \leq f(Y)/(Mg(Y))$, set $X = Y$ otherwise try again.
We note that $ P(U \leq f(Y)/(Mg(Y))) = 1/M$, so we would want $M$ as close to 1 as possible.

\noindent
{\it Proposition 6.1.} Let $f({\bf x}|\rho, \mbox{\boldmath$\mu$})$ and $g({\bf x}|\kappa, \mbox{\boldmath$\mu$})$ be the PKBD and  vMF distributions on $S^{d-1}$, respectively.
Given $\rho$ and $\mbox{\boldmath$\mu$}$,  $ f({\bf x}|\rho, \mbox{\boldmath$\mu$})<  M_\rho g({\bf x}|\kappa_\rho, \mbox{\boldmath$\mu$}),$
where
 $\kappa_\rho=\frac{d\rho}{1+\rho^2}$
 and \begin{equation} \label{eq:a4}
 M_\rho=(\frac{1}{c_d(\kappa_\rho) \omega_d \exp(\kappa_\rho )} )
 (\frac{1+\rho}{(1-\rho)^{d-1}} ). \end{equation} 
 
{\it Proof.} Given in the online supplementary materials.

\begin{table}
	\caption{Algorithm for generating a random variable from Poisson kernel-based density.}
\noindent\fbox{%
	\parbox{\textwidth}{\small
			\begin{spacing}{0.8}
			\begin{itemize}
				\item [1.] Generate Y from vMF density $g({\bf x}|\kappa_\rho, \mbox{\boldmath$\mu$})$  with 
				$\kappa_\rho=\frac{d\rho}{1+\rho^2}$,
				\item [2.] Generate $U \sim U(0,1)$ (independent of Y in Step 1),
				\item [3.] Let $M$ be as given in  (\ref{eq:a4}).
				If $U \leq f(Y)/(Mg(Y))$, return $X = Y$ ("accept") and stop;
				else go back to Step 1  ("reject") and try again.\\
			\end{itemize}
		\end{spacing}
			(Repeat steps 1 to 3 until acceptance finally occurs in Step 3).
	}
	}
\end{table}

\noindent
We note that, we can always use uniform distribution for the upper density but the efficiency, $1/M$, is much higher when using the vMF distribution. Table A6 in the supplementary material gives the efficiencies of the rejection method for simulating data from PKBD with a given concentration parameter $\rho$ using vMF and  uniform distribution as upper density, respectively.

\section{Practical Issues of Implementation of the Algorithm}

\noindent{\bf 1. Initialization Rule:}
To initialize the EM algorithm we randomly choose observation points as default initializers of the centroids.  This random starts strategy has a chance of not obtaining initial representatives from the underlying clusters. Therefore we choose as the final estimate of the parameters the one with the highest likelihood. 
Another approach that is commonly used for initialization is to use {\it K}-means to obtain the initial estimates of the centroids, where \textit{K}-means is initialized with multiple random starts. However, direct multiple random starts initialization performed as well as the more computationally expensive \textit{K}-means initialization and so we simply used the approach based on random starts. The initial values of all the concentration parameters for the components were set to 0.5 and we start with equal mixing proportions.

An alternative approach given in \citet{Duwairi} was used for initialization of centroids by \citet{Golzy}. However, the approach based on randomly selecting observations for initialization seems to provide a clustering solution with higher macro precision/recall, than the approach given in  \citet{Duwairi} in our context, particularly in the cases where the cluster centroids are close to each other.

\noindent{\bf 2. Stopping Rule Criteria:}
We use the following stopping rules: 
\begin{itemize}
\item either run the algorithm until the change in log-likelihood from one iteration to the next is less than a given threshold, or 
\item	run the algorithm until the membership is unchanged from one iteration to the next.
\end{itemize}
An alternative stopping rule is based on the maximum 
number of iterations needed to obtain "reasonable" results.

\noindent{\bf 3. Number of Clusters:}
An important problem in clustering is the estimation of the number of clusters and the literature includes a number of methods \citep{Rousseeuw, Tibshirani, Fraley02, Tibshirani05, Fujita}. The tables presented in the simulation section assume a known number of clusters. That is, the number of clusters is provided as input to the clustering algorithm.

We now briefly discuss a natural method for estimating the number of clusters when the model we use is mix-PKBD. The idea is simple, a determination on the number of clusters can be made on the basis of the first elbow that appears on the empirical densities distance plot, which we now define. 

The empirical densities distance plot depicts the value of the empirical  distance between the fitted mix-PKBD model $\hat G$ and $\hat F$, that is $ D_K(\hat F, \hat G)$ (y-axis), and the number of clusters $M$ (x-axis). \cite{Lindsay08} 
defined the quadratic distance between two probability measures by 
\[
D_K(F,G)=\int \int K({\bf x},{\bf y}) d(F-G)({\bf x})d(F-G)({\bf y})=\int \int K_{\rm cent(G)}({\bf x},{\bf y}) dF({\bf x})dF({\bf y}),\]
where $
K_{\rm cent(G)} (\textbf{x,y})=K(\textbf{x,y})- K(G,\textbf{y})- K(\textbf{x},G)+K(G,G) $,  
$K(G, {\bf y})=\int K({\bf x},{\bf y}) dG({\bf x})$.  $K(\textbf{x},G)$ is similarly defined, and $ K(G,G)=\int \int K({\bf x},{\bf y}) dG({\bf x})dG({\bf y}).$
The following algorithm is used to estimate the number of clusters when the model used is mix-PKBD.

\begin{table}[h!]
	\caption{Algorithm for computing the empirical densities distance plot for estimating the number of components.}
	\noindent\fbox{%
		\parbox{\textwidth}{\small
			\begin{spacing}{0.8}
				\begin{itemize}
					\item Run the clustering algorithm for different values of clusters M, for example  $M=1, 2, \cdots, 10$. 
					\item For each $M$, calculate the empirical distance between the fitted mixture model and the empirical density estimator. That is, compute
					
					\[ D_{K_\beta}(\hat F,\hat G_M)= 1/n^2 \sum_{i=1}^n  \sum_{j=1}^n  K_\beta ({\bf x_i},{\bf x_j}) -2/n \sum_{k=1}^M  \sum_{i=1}^n  {\hat \pi}_{k}  K_{\beta \hat \rho_k}({\bf x_i},\mbox{\boldmath$\hat \mu$}_k) + \sum_{k=1}^M   {\hat \pi}_{k}
					K_{\beta \hat \rho^2_k}(\mbox{\boldmath$\hat \mu$}_k,\mbox{\boldmath$\hat \mu$}_k)
					\]				
					
					\item Plot $D_{K_\beta}(\hat F,\hat G_M)$ versus $M$.
					
					\item  The location of a first elbow in the plot indicates the estimated number of clusters.
				\end{itemize}
			\end{spacing}
		}
	}
\end{table}
The plot of the number of clusters $M$ versus the distance 
$D_{K_\beta}(\hat F,\hat G_M)$  is called the {\it empirical densities distance plot} or simply {\it the distance plot} and its first elbow indicates the  estimated number of clusters.  Section B of Appendix A in the supplemental on-line material provides details on the calculation of $D_{K_\beta}(\hat F,\hat G_M)$. Here we note that the computation of the distance depends on a parameter $\beta$. Figures A1-A3 in the supplemental  material present the empirical distance plots as a function of various $\beta$ values and number of clusters.

To evaluate the performance of the proposed algorithm for estimating the number of clusters we performed a simulation study the  results of which are presented in section B of Appendix A (supplemental  material).  Briefly, we generate 50 replication samples on  the three-dimensional sphere according to the following specifications.  Each replication's sample size is 100; data are generated from a mixture of three equally weighted PKBD with mean vectors $(1,0,0), (0,1,0) $ and $(0,0,1)$ and the same concentration parameter $\rho$. We use a Poisson kernel with tuning parameter $\beta=0.1, 0.2$ and $0.5$ to calculate the  $D_{K_\beta}(\hat F,\hat G)$. The results, presented in Appendix A (see supplemental material, Table A1 and Figures A1-A3) indicate that, in general, the method works well. Specifically, when $\beta=0.1,0.2$ the method identifies the correct number of clusters for all $\rho \in [0.2,0.9]$. However, when $\beta=0.5$ the method identifies correctly the number of clusters for 
$\rho \in (0.4,0.9] $ and overfits when $\rho \in [0.2,0.4]$. Further work is needed to understand the impact of selecting $\beta$ on the estimation of the number of clusters.

\noindent{\bf 4. Robustness:}
\citet{Banfield} propose to model noise in the data by adding an additional mixture component in the model to account for the noise.  For directional data, it is natural to model the noise with a uniform distribution on the sphere.
Therefore for robustness analysis, we use the mixture model $
	f({\bf x}|\Theta)=\sum_{j=1}^M \alpha_j  f_j({\bf x}|\mbox{\boldmath$\theta$}_j)+ \frac{\alpha_0}{\omega_d}, $ $\alpha_j>0$ for all $j=0, \cdots M$, $\sum_{j=0}^M \alpha_j=1$.

The estimation method is the same as described in 3.2 with the difference that the pseudo posterior probabilities are defined by
\begin{equation}p(j|{\bf x_i},\Theta)=\left\{
\begin{array}{ll} \frac{ \alpha_0/\omega_d}{ f({\bf x_i}|\Theta) } & \mbox{if } j = 0 \\
\frac{ \alpha_j  f_j({\bf x_i}|\theta_j)}{ f({\bf x_i}|\Theta) }& \mbox{if } j=1, \cdots , M
\end{array},
\right.
\end{equation}
and we assign the points based on the following rule
\begin{equation} P({\bf x}_i, \Theta) := {\rm arg max}_{j\in\{0,1,2,...,M\}}\{p(j|{\bf x_i},\Theta)\}.\end{equation}

\section{Experimental Results}
\label{Sec:Exper}

In this section we present the results of several simulation studies that were designed to elucidate performance of our model in terms of a) imbalance in the mixing proportion; b) overlap among the components of the mixture densities; c) variety in the number of components and d) running time of the new algorithm in comparison with competing state-of-the-art clustering methods for directional data. These methods are mixtures of vMF distributions \citep{Ban05} and spherical \textit{k}-means \citep{Maitra}.
Performance is measured by macro-precision, macro-recall \citep{Modha} and also by the adjusted Rand index (ARI; \cite{Hubert}).

The statistical software R was used for all analyses. Spherical {\it K}-means clustering was performed by using the  function skmeans in R  \citep{Hornik12}.  Mixtures of vMF clustering was performed by using the function movMF in R  \citep{Horn}, and selecting the approximation given in \citet{Ban05} for estimation of the concentration parameters. The function adjustedRandIndex in R package "mclust" \citep{Fraley} was used to compute the adjusted Rand index.

\subsection{Simulation Study I:  Text Data}
The first set of simulations was based on text data. We simulated 100 Monte Carlo samples of text corpus using the Latent Dirichlet Allocation (LDA) model. 
The Latent Dirichlet allocation  model \citep{Blei03} postulates that documents are  represented as mixtures over latent independent topics, where each topic follows a multinomial distribution over a fixed vocabulary. Further, it uses the "bag of words" assumption, i.e. the order of the words in the document is immaterial, which guarantees exchangeability of random variables.

Thus, LDA assumes the following generative process for the $d^{\rm th}$ document in a corpus {\ D}.
\begin{itemize}
	\item [1.] Choose $N_d \sim$ Poisson ($\xi$), where $\xi$ is the average of the document sizes.
	\item [2.] Choose $\mbox{\boldmath$\theta$}_d \sim $ Dir($\mbox{\boldmath$\alpha$}$), where the parameter $\mbox{\boldmath$\alpha$}$ is a $k$-vector with components $\alpha_i>0$. 
	\item [3.] For each $i=1, \cdots, N_d$:
		\begin{itemize}
			\item [a.] Choose a topic ${\bf z}_i \sim$ Multinomial ($\mbox{\boldmath$\theta$}_d)$, 	${\bf z}_i^T = (z_i^1, \cdots,   z_i^k)$.
			\item [b.] Choose a word $w_i$ from $P({\bf w}_i|{\bf z}_i, B)$, a multinomial probability conditioned on the topic ${\bf z}_i$, where  $B$ is the $k \times v$ word probabilities matrix.
		\end{itemize}
\end{itemize}

The dimensionality $k$ of the Dirichlet distribution (and thus the dimensionality of the topic variables $z$) is assumed known and fixed. The word probabilities are parametrized by $k \times v$ matrix $B=(\beta_{ij})$ where $\beta_{i,j}= P( w^j=1| z^i=1)$.

Let $V$ denote the vocabulary used in any given text with size $v$, and let $\xi$ denote the average document size. Each realization of text data from a LDA model with $k=3$ topics is generated with the following specifications of the parameters. We take  $\alpha_i=1/3$, for each $i=1,2,3$, hence $\mbox{\boldmath$\alpha$}^T=(1/3,1/3,1/3)$ is the vector in the Dirichlet distribution used in step 2 above. Furthermore, the word probabilities matrix $B$ was taken to have rows $\mbox{\boldmath$\beta$}_i= (\beta_{i1}, \cdots, \beta_{iv}) \sim {\rm Dirichlet}(\mbox{\boldmath$\lambda$})$, where $\mbox{\boldmath$\lambda$}^T=(1/v, \cdots, 1/v)$,  $v$ is the vocabulary size. 
Therefore, the words in each document were generated as   $w_i \sim {\rm Multinomial} (B^t*z_i)$, where $z_i \sim {\rm Multinomial} (\mbox{\boldmath$\theta$}_d)$). 

We observed that the sparsity, that is the frequency of  zeros appearing as entries  in the vector space model, will increase as the ratio of $v/\xi$ increases. For example, if $v/\xi=0.25$ then sparsity is almost 0\%, if $v/\xi=1$ then sparsity is about 40\% and if $v/\xi=3$ then sparsity will increase to about 70\%.

\begin{table}[h!]{\footnotesize
		\caption {Macro-precision (M-P), macro-recall (M-R) and adjusted Rand index (ARI) for 100 Monte Carlo replications. The number of clusters equals 3.	} 
		\begin{center} 
			\begin{tabular}{|c|c|c|c|l|l l l|}
				\hline 
				N & $\xi$ & $v$ & $v/\xi$ & Eval.	&  mix-PKBD &  mix-vMF & Spkmeans\\ 
				\hline 
				&&  && M-P &  0.957 (0.02) & 0.934 (0.09) &	0.953  (0.02)	\\
				150	& 200 & 50& 0.25 & M-R &  0.954 (0.02) & 0.936 (0.06)	& 0.949 (0.02)	\\
				&&& & ARI & 0.866 (0.06) &	0.836 (0.12) &	0.851 (0.06) 	\\		
				\hline	
				&&	&& M-P &  0.962 (0.02) &	0.945 (0.07) &	0.960 (0.02)\\
				100	& 150 & 50& $0.3\overline3$ & M-R &  0.959 (0.02) &	0.944 (0.05) &	0.956 (0.02)	\\
				&&& & ARI&  0.883 (0.05) &	0.853 (0.08) &	0.878 (0.06)	\\ 	
				\hline
				&& & & M-P &  0.960 (0.02) &	0.952 (0.04) & 0.959 (0.02)\\
				100	& 200 & 75& 0.375 & M-R &  0.956 (0.02) &	0.950 (0.05) &	0.956 (0.02)\\
				&&& & ARI&  0.873 (0.06)&	0.863 (0.09) &	0.871 (0.06)\\			
				\hline	
				&& && M-P & 0.906 (0.03) &	0.895 (0.07) &	0.903 (0.03)	\\
				100	& 20 & 50& 2.5 & M-R &  0.901 (0.04) &	0.894 (0.06) &	0.898 (0.04)		\\
				&&& & ARI&  0.726 (0.09) & 	0.697 (0.13) &	0.718 (0.09)\\ 	
				\hline	
				&& & & M-P &  0.931 (0.07) &	0.822 (0.19) &	0.924 (0.10)\\
				50	& 200 & 50& 0.25& M-R &  0.931 (0.06) &	0.853 (0.13)&	0.930 (0.08)		\\
				&&& & ARI& 0.808  (0.12) &	0.729 (0.20) & 0.814 (0.14)\\			
				\hline	
				&& && M-P & 0.921 (0.04) & 0.906 (0.06) & 0.919 (0.04)	\\
				50	& 30 & 60& 2 & M-R & 0.918 (0.04) & 0.904 (0.06) & 0.917 (0.04)		\\
				&&& & ARI&  0.773 (0.11) & 0.761 (0.12) & 0.766 (0.11)\\ 	
				\hline	
				&& && M-P & 0.890 (0.07) &	0.890 (0.07) & 0.904 (0.04)\\
				50	& 15 & 75&  5& M-R & 0.890 (0.06) & 0.890 (0.05) &  0.902 (0.04)\\
				&&& & ARI& 0.710 (0.11) & 0.692 (0.12) &  0.728 (0.10)\\		
				\hline	
				&& && M-P &	0.897 (0.11) &  0.852 (0.15) & 0.927 (0.06)			\\
				40	& 100 & 20& 0.2 & M-R &  0.896 (0.08) & 0.867 (0.11) & 0.928 (0.05)					\\
				&&& & ARI& 0.728 (0.17) & 0.694 (0.19) & 0.790 (0.15) \\ 	
				\hline	
				&& && M-P &  0.877 (0.12) & 0.853 (0.15) & 0.905 (0.04)	\\
				40	& 30 & 60& 2 & M-R & 	0.887 (0.9) &  0.874 (0.10) & 0.906 (0.05)		\\
				&&& & ARI&  0.716 (0.15) & 0.696 (0.17) & 0.735 (0.11)\\ 	
				\hline				
			\end{tabular}	
		\end{center}}
	\end{table}

We compare the performance of the mixture of Poisson kernel-based distributions (mix-PKBD) with the state of the art mixture of vMF distributions (mix-vMF), and spherical {\it K}-means (Spkmeans) algorithm. 

Table 4 presents the mean of  macro-precision/recall and adjusted Rand index together with their associated standard deviations.  
The results indicate that when the sparsity of the data is low (i.e. 0.25, 0.33 or 0.375), and after taking into account the standard deviation, mix-PKBD outperforms mix-vMF, especially when $N=50$ with respect to all metrics involved. As the sparsity increases (i.e. $v/\xi$ has larger values) we see that the precision and recall of mix-vMF decreases, and its performance is the lowest among the three methods; mix-PKBD, in this case, performs slightly better than Spkmeans.
Note that, in the case where $\xi=15$, the vocabulary size $v$ is five times the average document size $\xi$, which produces a vector space model with very high percentage of sparsity.  

\subsection{Simulation Study II}

The goal of this set of simulations is to study the performance of the algorithm under a variety of conditions such as different sample sizes ($N$), dimensions ($d$), components in the mixture ($k$), distributions of the different components and proportion of the noise ($\pi_1$) in the data as expressed by a uniform distribution component incorporated in the mixture.

{\it Effect of proportion of noise data (uniform) on performance:} Figure 3 plots the different performance measures as a function of the mixing proportion $\pi_1$ of a uniform distribution on the $5$-dimensional sphere and a PKBD ($\rho=0.9$) distribution. The plots indicate that when the proportion of the uniform data gets large, mix-PKBD achieves the highest macro-recall, ARI and precision.

\begin{figure}[h!]
	\begin{center}
		\includegraphics[width=6.5in]{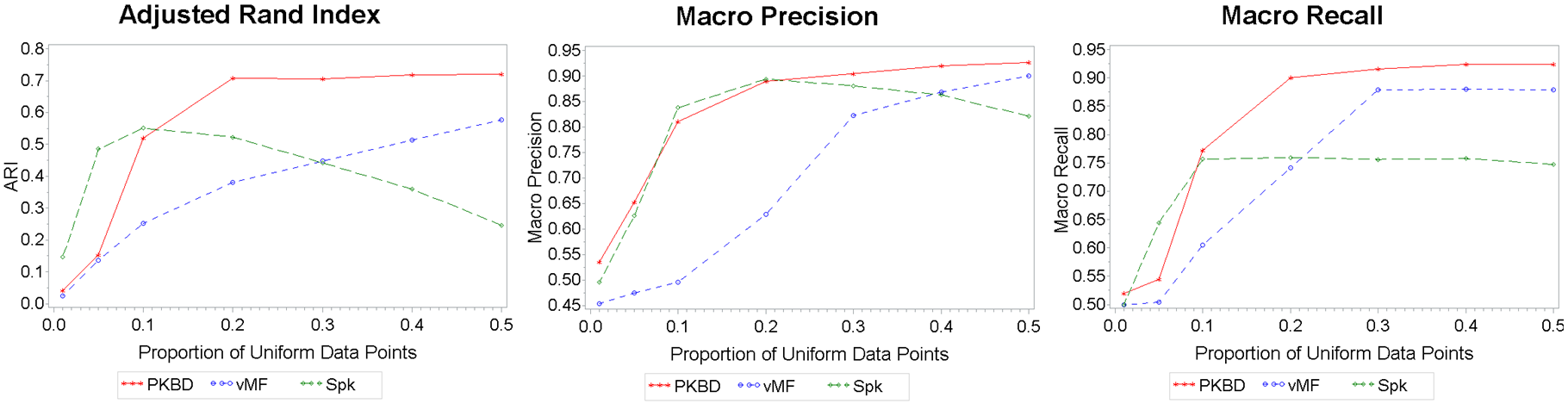}
	\end{center}
	\caption{\footnotesize ARI, Macro-Precision, Macro-Recall of mix-PKBD, mix-vMF and Spkmeans algorithms. Data are generated from a mixture of a uniform distribution with proportions given in the $x$-axis and a PKBD ($\rho=0.9$) distribution. Sample size is 200 and the number of Monte Carlo replications is 100. Dimension $d=5$. }
\end{figure}

{\it Effect of overlapping components:} We define overlap of components by how close their centers are on the scale of the cosine of the angle created by the vector of the individual centroids.  Specifically, we generated data first from a mixture of three component densities, one uniform and two PKBD ($\rho=0.9$), with sample size equal to $200$. The number of Monte Carlo replications is $100$. To be able to control the cosine of the angle between two centroid vectors, and therefore the component overlap, we consider the centroid vectors defined as $ \mbox{\boldmath$\mu$}_1^T=(1,0,0), \mbox{\boldmath$\mu$}_2^T=(a,0,\sqrt{1-a^2})$.  Then $\cos(\mbox{\boldmath$\mu$}_1,\mbox{\boldmath$\mu$}_2)=a$, and  the value of cosine between the two centers can be controlled as the parameter $a$ varies.

Figure 4 plots the ARI, macro precision and macro recall as a function of the cosine between the two centroids. Here, we study the effect of overlap in the presence of noise. Figure 4 shows that the new method outperforms in terms of ARI and macro recall the mix-vMF and Spkmeans algorithms and  it performs equivalently to mix-vMF in terms of macro precision when the cosine between the centroids is less than or equal to 0.3.

 \begin{figure}[h!]
 	\begin{center}
 		\includegraphics[width=6.5in]{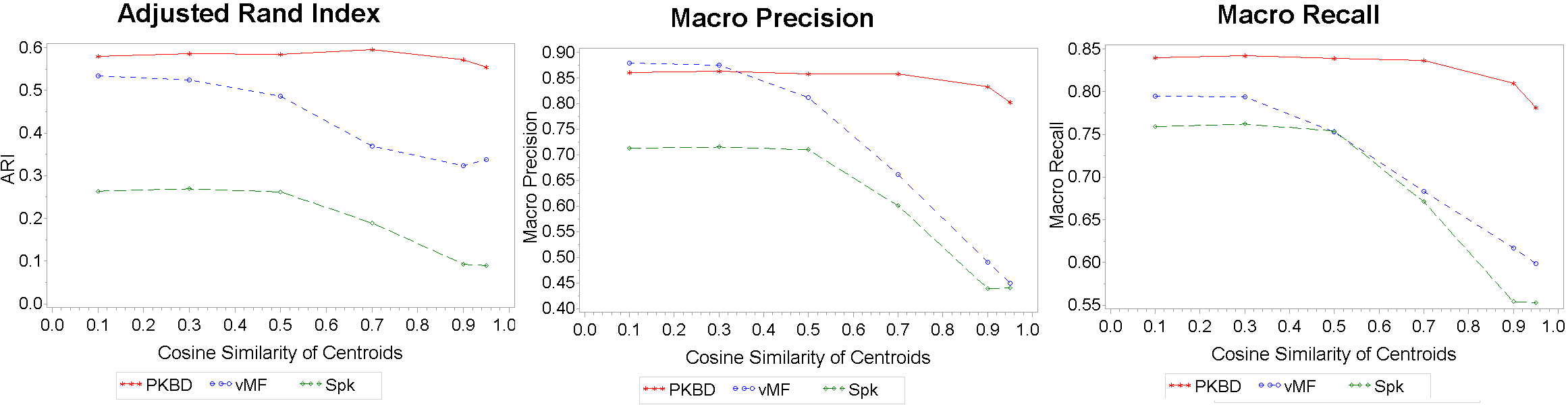}
 	\end{center}
 	\caption{\footnotesize ARI, Macro-Precision, Macro-Recall of mix-PKBD, mix-vMF and Spkmeans algorithms. Data are generated from a mixture of one uniform (50\%) and two equaly weighted PKBD ($\rho=0.9$) distributions with the cosine similarity of centroids given in the $x$-axis. Sample size is 200 and the number of Monte Carlo replications is 100. Dimension $d=3$. }
 \end{figure}

We then generated data on the $3$-dimensional sphere from a mixture of three equally weighted PKBD ($\rho=0.9$), with sample size and the number of Monte Carlo replications as above. To be able to control the cosine of the angle between the centroid vectors, and therefore the component overlap, we consider the centroid vectors defined as $ \mbox{\boldmath$\mu$}_1^T=(1/a,0,1), \mbox{\boldmath$\mu$}_2^T=(-1/2a,\sqrt{3}/2a,1),$ and $\mbox{\boldmath$\mu$}_3^T=(-1/2a,-\sqrt{3}/2a,1)$ after normalizing to length one.  Then $\cos(\mbox{\boldmath$\mu$}_i,\mbox{\boldmath$\mu$}_j)=\frac{2a^2-1}{2(a^2+1)}, i\neq j, i,j=1,2,3.$
Therefore, the value of the  cosine between any two of the three centers can be controlled as the parameter $a$ varies.

Figure 5 plots the ARI, macro-precision and macro-recall of the three algorithms as a function of the cosine of the angle between any two centroid vectors.  The graph indicates the following: a) when the cosine value is small, indicating a small amount of overlap between the different components, mix-PKBD and Spkmeans exhibit the highest values of macro-precision and recall. However, when the cosine of the angle is greater than or equal to $0.7$, mix-PKBD exhibits the best performance. 
 
\begin{figure}[h!]
	\begin{center}
		\includegraphics[width=6.5in]{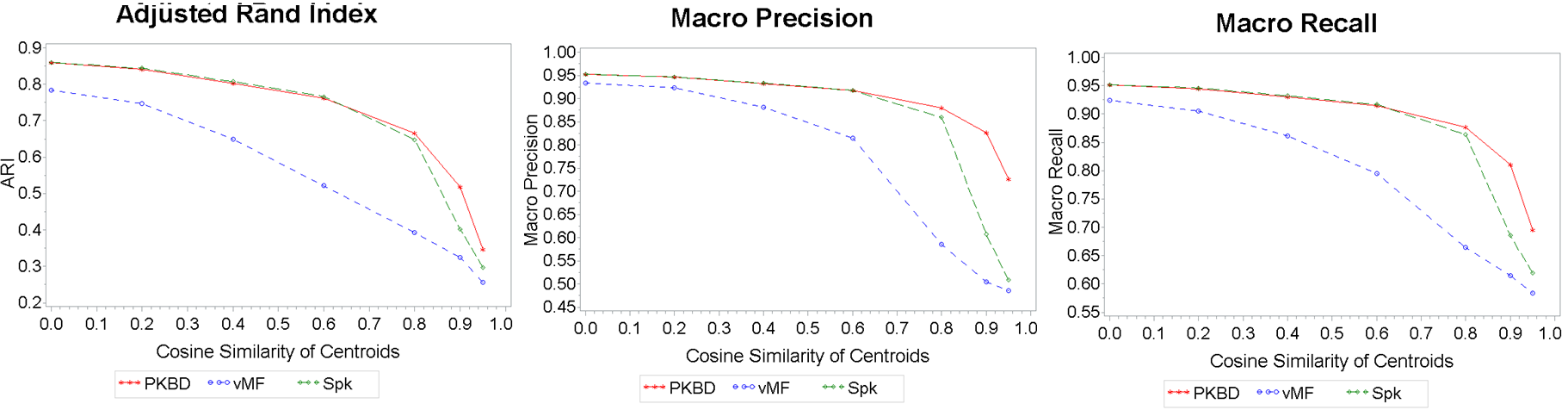}
	\end{center}
	\caption{\footnotesize ARI, Macro-Precision, Macro-Recall of mix-PKBD, mix-vMF and Spkmeans algorithms. Data are generated from a mixture of three equaly weighted PKBD ($\rho=0.9$) distributions with the cosine similarity of centroids given in the $x$-axis. Sample size is 200 and the number of Monte Carlo replications is 100. Dimension $d=3$. }
\end{figure}

Tables A2 and A3 of the supplemental material present ARI, macro-precision and macro recall of the three algorithms under consideration when the sample size increases but the dimension stays fixed, and when the dimension increases but the sample size stays fixed. Data of equal proportions were generated from a mixture of uniform and either PKBD or vMF densities. Overall, when the sample size increases mix-PKBD  seems to have the highest macro-precision, recall and ARI, after taking into account the standard error of the estimates of the performance measures. When the sample size is fixed but the dimension increases, mix-PKBD performs almost equivalently with mix-vMF algorithm, while Spkmeans indicates lesser performance than the other two algorithms.\\

Figures A4-A6 of the supplemental material investigate the effect of number of clusters, and also the effect of the value of the concentration parameter of the PKBD components on the performance of the mix-PKBD algorithm, while figure A7 shows that there is no significant difference in the run time between  the different algorithms.

\subsection{Application to Real Data}

We now apply our method on well known data sets; detailed description of the data sets is provided in the on-line supplemental material. The data points are projected onto the sphere by normalizing them so the associated vectors have length one.  
The data sets were selected to exhibit different sample sizes, dimensions, and number of clusters. For the text data sets, we used Correlated Topic Modeling (CTM) \citep{Blei07, Grun} for the dimension reduction and topics were used as features instead of  words. 
 
Table A4 of the supplemental material  presents the results for all examples, that show, in most cases, mix-PKBD exhibits higher values of the evaluation indices than mix-vMF oe Spkmeans. To further illustrate the methods, we  discuss here in some detail the Seeds and the Crabs data sets.
	
\noindent
{\it Seeds Data:} We fitted a mix-PKBD$(\rho_i)$ model to this data set. The empirical densities distance plot $(\beta=0.1)$ estimated the number of clusters to be 3 (see Figure A9 in Appendix A).  The mixing proportions are $(0.2578, 0.3302, 0.4120)$ and  the concentration parameters of the PKBD densities were 0.9922, 0.9866 and 0.9866, respectively. The inner products $\mbox{\boldmath$\mu$}_1 . \mbox{\boldmath$\mu$}_2 $ ,  $\mbox{\boldmath$\mu$}_1 . \mbox{\boldmath$\mu$}_3$ and $\mbox{\boldmath$\mu$}_2 .  \mbox{\boldmath$\mu$}_3$ where $\mbox{\boldmath$\mu$}_1, \mbox{\boldmath$\mu$}_2, \mbox{\boldmath$\mu$}_3$ are the cluster centroids are 0.9839, 0.9974 and 0.9916 indicating that the three clusters have a fair amount of overlap. Figure A10 indicates graphically  the overlap among the different clusters.

\noindent
{\it Crabs Data:} The second data set is the crabs data, details of which are presented in the supplemental material (Section C of Appendix A). For this data set we first run mix-PKBD with the number of clusters equal to 2. We also run mix-vMF and Spkmeans again using two clusters with the two color species indicating the classes.  In this case, the performance of mix-VMF and Spkmeans was surprisingly poor.  To assess cluster homogeneity we present Figures A11 and A12 (supplemental material), and  the scatter plot matrices for this data set by species (blue or green crabs) and by sex, respectively.  Each clustering algorithm discovers structure in the data; the mix-PKBD model seems to cluster the data according to species where the degree of separation is higher than clustering according to sex. The clusters produced by mix-vMF and Spkmeans are more likely to correspond to clustering by gender and not species.  

We also computed the empirical densities distance plot (see Figure A13 of the supplemental material). This plot estimates the number  of clusters as four and it seems that clusters are formed by species and gender. We also  run mix-vMF and Spkmeans models with four clusters. Table S6 of the supplemental material, Appendix B presents the performance measures for all models, indicating that all perform equivalently. 

\section{Discussion \& Conclusion}
\label{Sec:Conc}
 We introduced and discussed a novel model for clustering directional data that is based on the Poisson kernel.  We presented connection of the Poisson kernel based density function with other models that are used for the analysis of directional data.  We developed a clustering algorithm that is based on a mixture of PKBD, studied the identifiability of the proposed model, the convergence of the associated algorithm and, via simulation and application to real data, we compared the performance of  the proposed clustering algorithm with the algorithm proposed by \cite{Ban05} and Spkmeans. Furthermore, we investigated practical issues associated with the operationalization of our procedure, and proposed a natural method to estimate the number of clusters from the data.
 
 Our methods are based on mixtures of PKBD and as such are model based. \cite{McNi} argues in favor for model based clustering methods. Our results indicate that our methods, in all cases examined, exhibit excellent performance when compared with state of the art methods.
 
 An interesting aspect of clustering based on the mix-PKBD model is the robustness exhibited in the presence of noise. 
Our model exhibits the best performance in terms of
macro-precision and recall especially when the proportion of noise is high. On the other
hand, mix-PKBD performs similarly with the other two methods when the amount of
noise is low. There are cases where mix-PKBD has inferior performance than mix-vMF
and Spkmeans in terms of macro-precision and recall. We generated data from a mixture of
vMF($\kappa = 40$) and a PKBD($\rho = 0.8$) distributions. The cosine between the center vectors of the components was 0.75 indicating an approximately $41\circ$ angle. When the mixing propor
tion of the PKBD(0.8) was greater than 0.6, mix-PKBD exhibited higher macro-precision
\& recall than mix-vMF and Spkmeans (data are not shown). Note that PKBD(0.8) was
selected so that the mode of vMF and PKBD(0.8) distributions are approximately the
same. The dimension of the data in this case equals three.

Poisson kernel-based mixture models oﬀer a natural way to estimate the number of
clusters. We introduced the empirical densities distance plot that can be used to estimate
25
the number of clusters when the data are clustered using mix-PKBD. We note here that the
empirical densities distance plot depends on a tuning parameter $\beta$. When the clustering
model is a mixture of PKBD with a common $\rho$, we conjecture that $\beta$ can be selected such that $\beta \leq \hat \rho$, where $\hat \rho$  is an estimate of $\rho$. When the clustering model is a mixture of PKBD with diﬀerent parameters ρi, we conjecture that $\beta \leq 
\min\limits_i \{ \hat \rho_i: 1\leq i\leq M\}$. Additional work is needed to fully understand the selection of the tuning parameter β and the performance
of the distance plot.

\bigskip
\begin{center}
	{\large\bf SUPPLEMENTAL MATERIALS}
\end{center}

%\begin{spacing}{1}
\begin{description}
	
\item[Title:] Appendix A "Poisson Kernel-Based Clustering on the Sphere: Convergence Properties, Identifiability, and a Method of Sampling"\\	
\noindent 
Appendix A is organized in four sections. Section A presents detailed proofs of the propositions that appear in this manuscript. Section B presents calculations and simulations associated with the estimation of the number of clusters. 
Section C includes additional simulation examples and application of our methods in a variety of data sets, while Section D offers additional tables and graphs illustrating further comparison of the PKBD model with the von Mises-Fisher model, and the elliptically symmetric angular Gaussian (ESAG) model.
\item [Title:] Appendix B; Codes Associated with "Poisson Kernel-Based Clustering on the Sphere: Convergence Properties, Identifiability, and a Method of Sampling"
	
\end{description}

\spacingset{1.5}

%\end{spacing}

\end{document}

\def\spacingset#1{\renewcommand{\baselinestretch}%
	{#1}\small\normalsize} \spacingset{1.5}

\if0\blind
{
	\title{\Large \bf Appendix A "Poisson-Kernel Based Clustering on the Sphere: Convergence Properties, Identifiability, and a Method of Sampling"}
	\author{Mojgan Golzy and 
		Marianthi Markatou$^*$  \\
		Department of Biostatistics, University at Buffalo, Buffalo, NY
	}
	\maketitle
} \fi

\if1\blind
{
	\bigskip
	\bigskip
	\bigskip
	\title{ \Large \bf Supplementary Material for "Poisson-Kernel Based Clustering on the Sphere: Convergence Properties, Initialization Rules and a Method of Sampling"}
	\medskip
} \fi

{\bf \Large Section A: Detailed Proofs}\\

\noindent
{\bf \large Proofs of the Propositions in Section 3}\\
	
	\noindent
{\it Proof of Proposition 3.1.} The density of ${\bf u}$ can be written as
	\[
	f({\bf u}|\mbox{\boldmath$\mu$}=0, \Sigma)= \frac{1}
	{\omega_d  \sigma \{ (1/\sigma^2)u_1^2+\sum_{i=2}^{d} u_i^2 \}^{d/2}}\\
	=\frac{1}
	{\omega_d  \sigma \{ (1/\sigma^2)u_1^2+1-u_1^2\}^{d/2}}.\\
	\]
	Let $u_1=\cos (\theta)$ then $2u_1^2=1+\cos(2\theta)$ and so
	\[
	f({\bf u}|\mbox{\boldmath$\mu$}=0, \Sigma)= \frac{1}
	{\omega_d  \sigma \{ \frac{\sigma^2+1}{2 \sigma^2} -\frac{\sigma^2-1}{2 \sigma^2}\cos(2\theta)\}^{d/2}}.\\  
	\]
	For any constant $c>0$, 
	\[
	f({\bf u}|\mbox{\boldmath$\mu$}=0, \Sigma)= \frac{c^d/\sigma}{\omega_d  \{ c^2\frac{\sigma^2+1}{2 \sigma^2} -c^2\frac{\sigma^2-1}{2 \sigma^2}\cos(2\theta)\}^{d/2}}.
	\]
	
	Suppose $R^d_\theta {\bf u}$ has Poison distribution with parameters $\rho$ and $ {\bf y}$ then $R_\theta {\bf u}$ has a density function  given by
	\[  \frac{1-\rho^2}{\omega_d \{1+\rho^2 -2 \rho (R^d_\theta {\bf u}){\bf \cdot}  {\bf y}\}^{d/2}}. \]
	
	If $\sigma^2<1$ then  we let  ${\bf y}=(-1,0, \cdots, 0)$ and so  $(R^d_\theta {\bf u}){\bf \cdot}  {\bf y}=-\cos(2\theta)$ and if  $\sigma^2>1$ then we let  ${\bf y}=(1,0, \cdots, 0)$ and so  $(R^d_\theta {\bf u}){\bf \cdot}   {\bf y}=\cos(2\theta)$.  So, the two densities are equal if there is a constant $c$ such that the following system of equations have a solution.\\
	\[1-\rho^2=c^d/\sigma,\]
	
	\[1+\rho^2=c^2\frac{\sigma^2+1}{2 \sigma^2},\]
	
	\[2 \rho =c^2\frac{|\sigma^2-1|}{2 \sigma^2}.\]
	
	If $d=2$ then this system of equations has a unique solution $c=\frac{2 \sigma}{\sigma+1}$,  $\rho=\frac{\sigma-1}{\sigma+1}$. But for $d>2$ this system of equations has no solution and so $R^d_\theta {\bf u}$ has Poisson kernel-based distribution if and only if d=2. \\

\noindent
{\it Proof of Proposition 3.2.} To obtain the estimates of the parameters we maximize the Lagrangian for the second term of the complete likelihood expression, given by
{\small 	\begin{equation}
	\sum_{j=1}^M\sum_{i=1}^N   \{ ln(1-\rho_j^2) - ln(\omega_d)- d \  ln \big \|{\bf {\bf x_i}}-\rho_j \mbox{\boldmath$\mu_j$} \big \|\} \times
	p(j|{\bf x_i}, \Theta^{(t-1)})
	+\sum_{j=1}^M \lambda_j (1-\big \|\mbox{\boldmath$\mu$}_j\big \|^2),	\end{equation}}
subject to $0<\rho<1$ for each $j$.
		Differentiating the Lagrangian with respect to $\rho_k, \mbox{\boldmath$\mu_k$} $ and $\lambda_k$ we obtain
		\begin{equation} \label{eq:rho1}
			\partial l/ \partial \rho_k= \frac{-2 \rho_k}{1-\rho_k^2} \sum_{i=1}^n p(k|{\bf x_i}, \Theta^{(t-1)}) +d \sum_{i=1}^n \frac{({\bf {\bf x_i}}.\mbox{\boldmath$\mu_k$}-\rho_k  )}{\big \|{\bf {\bf x_i}}-\rho_k \mbox{\boldmath$\mu_k$} \big \|^2 }p(k|{\bf x_i}, \Theta^{(t-1)}),
		\end{equation}
		
		\begin{equation}\label{eq:mu1}
			\partial l/ \partial \mbox{\boldmath$\mu_k$} = d \rho_k\sum_{i=1}^n \frac{({\bf x_i}-\rho_k \mbox{\boldmath$\mu_k$} )}{\big \|{\bf x_i}-\rho_k \mbox{\boldmath$\mu_k$}\big \|^2}p(k|{\bf x_i}, \Theta^{(t-1)}) -2 \lambda_k \mbox{\boldmath$\mu_k$}, 
		\end{equation}
		
		\begin{equation}\label{eq:lambda}
			\partial l/ \partial \lambda_k=1-\big \|\mbox{\boldmath$\mu_k$} \big \|^2 .
		\end{equation}
		
		For convenience of the mathematical analyses, we use a variant of the EM algorithm by using the old estimates of the parameters ($\rho_k^{(t-1)}$ and $\mbox{\boldmath$\mu_k$}^{(t-1)}$) in the denominators of the equations (\ref{eq:rho1}) and (\ref{eq:mu1}) and use notation $w_{ik}$ for $\frac{p(k|{\bf x_i}, \Theta)}{\big \|{\bf x_i}-\rho_k\mbox{\boldmath$\mu_k$} \big \|^2}$. 
		Then equations (\ref{eq:rho1}) and  (\ref{eq:mu1}) can be rewritten as
		\begin{equation} \label{eq:rho3}
			\partial l/ \partial \rho_k= \frac{-2\rho_k }{1-\rho_k^2} (n \alpha_k^{(t)})+d \sum_{i=1}^n w_{ik}^{(t-1)} {\bf x}_i {\bf \cdot}\mbox{\boldmath$\mu_k$} - d \rho_k\sum_{i=1}^n w_{ik}^{(t-1)},
		\end{equation}
		
		\begin{equation}\label{eq:mu3}
			\partial l/ \partial \mbox{\boldmath$\mu_k$} = d \rho_k \sum_{i=1}^n w_{ik}^{(t-1)} {\bf x}_i -d \rho_k^2 \sum_{i=1}^n w_{ik}^{(t-1)} \mbox{\boldmath$\mu_k$}  -2 \lambda_k \mbox{\boldmath$\mu_k$}.
		\end{equation}
		Setting equations  (\ref{eq:lambda}) and (\ref{eq:mu3}) equal zero we get two solutions for $\mbox{\boldmath$\mu_k$}$,
		\begin{equation} \mbox{\boldmath$\mu_k$} =  \frac{\sum_{i=1}^n w_{ik}^{(t-1)} {\bf x_i}}{\big \|\sum_{i=1}^n w_{ik}^{(t-1)} {\bf x_i}\big \|} {\rm ~~~~~ and ~~~~~}  \mbox{\boldmath$\mu_k$} = - \frac{\sum_{i=1}^n w_{ik}^{(t-1)} {\bf x_i}}{\big \|\sum_{i=1}^n w_{ik}^{(t-1)} {\bf x_i}\big \|}.\end{equation}
		We note that if we start with an initial estimate of  $ \mbox{\boldmath$\mu_k$}$ in the same direction as the true value then the dot product $\mbox{\boldmath$\mu_k$}^{(t-1)}.\mbox{\boldmath$\mu_k$}^{(t)}$ should be positive, at each iteration. 
		Therefore, if we start with a good initial estimate we have
		\begin{equation}\label{eq:mu}\mbox{\boldmath$\mu_k$}^{(t)} =  \frac{\sum_{i=1}^n w_{ik}^{(t-1)} {\bf x_i}}{\big \|\sum_{i=1}^n w_{ik}^{(t-1)} {\bf x_i}\big \|},\end{equation}
		and (from  \ref{eq:lambda} and \ref{eq:mu3})
		\begin{equation}\label{eq:lamda2} d \rho_k \big \|\sum_{i=1}^n w_{ik}^{(t-1)} {\bf x_i}\big \|=d \rho_k^2 \sum_{i=1}^n w_{ik}^{(t-1)} +2 \lambda_k. \end{equation}
		Using $\mbox{\boldmath$\mu_k$}^{(t)}$ in equation (\ref{eq:rho3}) and setting it equal zero, we have
		\begin{equation} \label{eq:rho4}
			\frac{-2n \rho_k \alpha_k^{(t)}}{1-\rho_k^2} ~ + d \big \|\sum_{i=1}^n w_{ik}^{(t-1)} {\bf x_i}\big \| - d \rho_k \sum_{i=1}^n w_{ik}^{(t-1)}=0.
		\end{equation}
		We note that, if $\rho_k=0$ then the left hand side of  (\ref{eq:rho4}) is positive and is negative if $\rho_k \rightarrow 1$. Therefore this equation has a solution between $0$ and $1$.
		
		Hence the estimates of the parameters   ${\mbox{\boldmath$\mu_k$}}$ and $\rho_k$ can be calculated using the following iterative re-weighted algorithm; Let $ \Theta^{(0)}=\{ \alpha^{(0)}_1,\cdots, \alpha^{(0)}_M, \rho^{(0)}_1,\cdots, \rho^{(0)}_M, {\mbox{\boldmath$\mu$}}^{(0)}_1, \cdots, { \mbox{\boldmath$\mu$}}^{(0)}_M\} $ be the initial values of the parameters, then we define ${w}_{ik}^{(t-1)}, {\alpha_k}^{(t)}, {\rho_k}^{(t)}$ and $ {\mbox{\boldmath$\mu$}}_k^{(t)}$ for $ t=1, 2, \cdots $ iteratively as follow;
		\begin{equation} \label{eq:estimates}
			\begin{array}{ll}
				{w}_{ik}^{(t-1)} &=\frac{p(k|{\bf x_i}, \Theta^{(t-1)})}{\big \|{\bf x_i}-{\rho_k}^{(t-1)} \mu_k^{(t-1)}\big \|^2}, \\
				{\alpha_k}^{(t)} &=(1/N) \sum_{i=1}^N p(k|{\bf x_i}, \Theta^{(t-1)}),\\
				{\mbox{\boldmath$\mu$}}_k^{(t)} & =  \frac{\sum_{i=1}^n {w}_{ik}^{(t-1)} {\bf x_i}}{\big \|\sum_{i=1}^n {w}_{ik}^{(t-1)} {\bf x_i}\big \|},  \\
				\rho^{(t)}_k &=\rho^{(t-1)}_k- \frac{ g_k(\rho^{(t-1)}_k)}{ g_k'(\rho^{(t-1)}_k)},
			\end{array}
		\end{equation}
		where
		$ g_k(y)=\frac{-2n y \alpha_k^{(t-1)}}{1-y^2}+ d  \big \|\sum_{i=1}^n w^{(t-1)}_{ik}{\bf x}_i\big \| -d y \sum_{i=1}^n  w^{(t-1)}_{ik},$  and $ g_k'$ is derivative of $g_k$. \\

{\bf \large Proofs of the Propositions in Section 4}\\

In order to investigate the linear independence of the Poisson kernel-based densities, we need some basic results, which are given here.  The $d$-variate Poisson kernel-based distribution $ K_\rho(., \mbox{\boldmath$\mu$}) $ can be written as 
\begin{equation}\label{eq:14}	
	K_\rho ({\bf x},\mbox{\boldmath$\mu$})=\frac{1}{\omega_d}\sum_{n=0}^\infty \rho^n Z_n({\bf x},\mbox{\boldmath$\mu$}), \end{equation} 
where $Z_n({\bf x},\mbox{\boldmath$\mu$})$  is called the zonal harmonic of degree $n$ with
pole $\mbox{\boldmath$\mu$}$, and satisfies the following equation \citep{Dai}.
For every $\mbox{\boldmath$\xi$}, \mbox{\boldmath$\eta$} \in S^{d-1}$, 
\begin{equation} \label{eq:15} \frac{1}{\omega_d} \int_{S^{d-1}} Z_m(\mbox{\boldmath$\xi$},{\bf y}) Z_n(\mbox{\boldmath$\eta$},{\bf y})d\sigma({\bf y})=Z_n(\mbox{\boldmath$\xi$},\mbox{\boldmath$\eta$})	\delta_{n,m}.	
\end{equation}

\noindent
{\it Proof of Lemma 4.2.} Let $ g(x)=\sum_{j=1}^K \alpha_j K_{\rho_j}(x, \mu_j)=0$. 
Then for each $n$, 
\[ \begin{array}{ll}
0 &= (1/\omega_d) \int_{S^{d-1}}g(y) Z_n(x,y)d \sigma(y)\\
&= \sum_{j=1}^K (\alpha_j /\omega_d) \int_{S^{d-1}} K_{\rho_j}(y, \mu_j) Z_n(x,y)d \sigma(y)\\
&= \sum_{j=1}^K (\alpha_j /\omega_d) \int_{S^{d-1}} \sum_{m=0}^\infty \rho_j^m Z_m(y,\mu_j)Z_n(x,y)d \sigma(y)\\
&= \sum_{j=1}^K \alpha_j \sum_{m=0}^\infty \rho_j^m \{ (1/\omega_d)\int_{S^{d-1}}Z_m(y,\mu_j)Z_n(x,y)d \sigma(y)\}\\
&= \sum_{j=1}^K \alpha_j \sum_{m=0}^\infty \rho_j^m Z_m(x,\mu_j) \delta_{n,m} \\
&= \sum_{j=1}^K \alpha_j  \rho_j^n Z_n(x,\mu_j)  \\
\end{array}\]
Therefore,   $ g(x)=0$ implies $ \sum_{j=1}^K \alpha_j \rho_j^n Z_n(x,\mu_j) =0$, for each $x$ and $n$. \\

 We recall from \cite{Dai} that, for each $ x, y \in S^{d-1}$, $d \geq 3$,
\begin{equation} \label{eq:zonal} |Z_n(x,y)|\leq |Z_n(x,x)|=dim {\cal H}_n^d,\end{equation}
where ${\cal H}_n^d $ is the linear space of real harmonic polynomials, homogeneous of degree $n$. This relationship will be used in the following proposition.\\

\noindent 
{\it Proof of Proposition 4.3.}  Let $ \sum_{j=1}^K \alpha_j K_{\rho_j}(x, \mu_j)=0$ for each $x$. By lemma 4.2, \\
$ \sum_{j=1}^K \alpha_j \rho_j^n Z_n(x,\mu_j) =0$, for each $x$ and $n$. Assume that there exists at least one $j$ so that $\alpha_j\neq 0$ and define
\begin{equation} j*:=\arg \max _{j=1, \cdots K}\{ \rho_j | \alpha_j\neq0\},\end{equation}
and so, for each $j\neq j*$
\begin{equation} lim_{n\rightarrow \infty} (\alpha_j/\alpha_{j*})(\frac{\rho_j}{\rho_{j*}})^n =0.\end{equation}
We choose $\epsilon $ small enough such that  $0< \frac{\epsilon}{K-1}<\frac{1}{K-1}$. For each $j\neq j*$, there exists a $N_j$ such that for each $m > N_j$
\begin{equation}|(\alpha_j/\alpha_{j*})(\frac{\rho_j}{\rho_{j*}})^m|<\frac{\epsilon}{K-1},\end{equation}
or equivalently, for each $m > N_j$ 
\begin{equation}|\alpha_j \rho_j^m|< \frac{\epsilon}{K-1} |\alpha_{j*}\rho_{j*}^m|.\end{equation}
Take $N_0= \max\{N_j: j\neq j*\}$ then,
\[|\alpha_j \rho_j^m|< \frac{\epsilon}{K-1} |\alpha_{j*}\rho_{j*}^m|, \hspace{0.5cm} ~{\rm for ~ each~ } j\neq j*  ~{\rm for ~ each~} m > N_0.\]

Setting $x= \mu_{j*}$, we get $ \sum_{j=1}^K \alpha_j \rho_j^n Z_n(\mu_{j*},\mu_j) =0$, for each $n$.  By (\ref{eq:zonal}),  $ |Z_n(\mu_{j*},\mu_j)|\leq |Z_n(\mu_{j*},\mu_{j*})|=dim {\cal H}_n^d$ and so
\[|\alpha_j \rho_j^m||Z_m(\mu_{j*},\mu_j)|< \frac{\epsilon}{K-1} |\alpha_{j*}\rho_{j*}^m||Z_m(\mu_{j*},\mu_{j*})|, \hspace{0.5cm} ~{\rm for ~ each~ } j\neq j*  ~{\rm for ~ each~} m > N_0.\]
Therefore,

\[
\begin{array}{ll}
0=|\sum_{j=1}^K \alpha_j \rho_j^m Z_m(\mu_{j*},\mu_j) | 
& \geq 
|\alpha_{j*} \rho_{j*}^m Z_m(\mu_{j*},\mu_{j*})|-\sum_{j\neq j*} |\alpha_j \rho_j^m Z_m(\mu_{j*},\mu_j) |\\
& > 
|\alpha_{j*} \rho_{j*}^m Z_m(\mu_{j*},\mu_{j*})|-\sum_{j\neq j*} \frac{\epsilon}{K-1} |\alpha_{j*}\rho_{j*}^m||Z_m(\mu_{j*},\mu_{j*})|\\
& =
|\alpha_{j*} \rho_{j*}^m Z_m(\mu_{j*},\mu_{j*})| \{1-\sum_{j\neq j*} \frac{\epsilon}{K-1} \} \\
& =
|\alpha_{j*} \rho_{j*}^m Z_m(\mu_{j*},\mu_{j*})| \{1-\epsilon\} \\
& =
\underbrace{|\alpha_{j*} \rho_{j*}^m ||\dim {\cal H}_m^d|}_{\neq 0} \underbrace{\{1-\epsilon\}}_{>0} \\
&>0,
\end{array}
\]
which is a contradiction.\\

\noindent{\bf \large Proofs of the Theorems in Section 5}\\
	
\noindent
	{\it Proof of Theorem 5.1.}
	For the proof of the first item we refer to \citet{Xu}. To prove the second item, we note that, 
	\begin{equation}\label{eq:mu3}
		\partial l/ \partial \mbox{\boldmath$\mu_k$} = d \rho_k \sum_{i=1}^n w_{ik}^{(t-1)} {\bf x}_i -d \rho_k^2 \sum_{i=1}^n w_{ik}^{(t-1)} \mbox{\boldmath$\mu_k$}  -2 \lambda_k \mbox{\boldmath$\mu_k$}.
	\end{equation}
Then
	\begin{equation}\label{eq:lamda2} d \rho_k \big \|\sum_{i=1}^n w_{ik}^{(t-1)} {\bf x_i}\big \|=d \rho_k^2 \sum_{i=1}^n w_{ik}^{(t-1)} +2 \lambda_k. \end{equation}
	implies
	\begin{equation}
		\partial l/ \partial \mbox{\boldmath$\mu_k$} = d \rho_k \sum_{i=1}^n w_{ik}^{(t-1)} {\bf x}_i - d \rho_k\big \|\sum_{i=1}^n w_{ik}^{(t-1)} {\bf x_i}\big \| \mbox{\boldmath$\mu_k$},\end{equation}
	and so
	\begin{equation} a_k^{(t-1)} \partial l/ \partial \mbox{\boldmath$\mu_k$}|_{\mbox{\boldmath$\theta_k$}=\mbox{\boldmath$\theta_k$}^{(t-1)}} =\frac{ \sum_{i=1}^n w_{ik}^{(t-1)} {\bf x}_i }{\big \|\sum_{i=1}^n w_{ik}^{(t-1)} {\bf x_i}\big \| }- \mbox{\boldmath$\mu^{(t-1)}_k$}=\mbox{\boldmath$\mu^{(t)}_k$}- \mbox{\boldmath$\mu^{(t-1)}_k$}.\end{equation}
	Therefore, ${\cal M}^{(t)}- {\cal M}^{(t-1)}= {\cal P}_{\cal M}^{(t-1)}\frac {\partial l}{\partial {\cal M} }|_{{\cal M}= {\cal M}^{(t-1)}}.$
	
	To prove the third item, we note that $ \alpha_k^{(t)}=1/n \sum_{i=1}^N p(k|{\bf x_i},\Theta^{(t-1)})$ and\\
	$ \sum_{i=1}^n w_{ik}^{(t-1)} {\bf x}_i \mbox{\boldmath$\mu_k$}^{(t)}=\big \|\sum_{i=1}^n w^{(t-1)}_{ik}{\bf x}_i\big \| ,$  and so
	$ g_k(\rho^{(t-1)}_k)=\partial l/ \partial \rho_k |_{\rho_k=\rho_k^{(t-1)}}. $
	Therefore,
	\begin{equation} {\cal R}^{(t)}-{\cal R}^{(t-1)}={\cal P}_{\cal R}^{(t-1)} \ \partial l/ \partial {\cal R}|_{{\cal R}={\cal R}^{(t-1)}}. \end{equation} \\

	\noindent{\it Proof of Theorem 5.2.}
	Let $\Theta=({\cal A, R,M})$ and $ {\cal P} (\Theta)= {\rm diag}({\cal P}_{\cal A},  {\cal P}_{\cal R},{\cal P}_{\cal M})$, we can combine the three items in the previous theorem as a single equation:
	
	\begin{equation} \Theta^{(t)}=\Theta^{(t-1)} +  {\cal P}(\Theta^{(t-1)})  \partial l/ \partial \Theta|_{\Theta=\Theta^{(t-1)}}. \end{equation}
	\citet{Xu} have shown that ${\cal P}_{\cal A}^{(t-1)}$ is a positive definite matrix. 
	${\cal P}_{\cal M}^{(t-1)}$ is a positive definite matrix, since $a_k^{(t-1)}>0$ for all $k$, and ${\cal P}_{\cal R}^{(t-1)}$ is a positive definite matrix, since 
	\begin{equation}- g'_k(y)= \frac{2n (1+y^2) \alpha^{(t-1)}_k}{(1-y^2)^2} + d  \sum_{i=1}^n w_{ik}^{(t-1)}>0,\end{equation}
	for all $y, k$. Therefore ${\cal P}(\Theta^{(t-1)}) $ is a positive definite matrix.
	Thus, the likelihood is guaranteed not to decrease after each iteration. Since $f({\bf x}|\Theta)$ is bounded, the log-likelihood function $l$ is bounded, and so, it is guaranteed to converge to a local maximum. \\

\noindent{\bf \large Proof of the Proposition in Section 6}\\

\noindent
{\it Proof Proposition 6.1.} Let $h(t)=\log \frac{1-\rho^2}{ (1+\rho^2-2\rho t)^{d/2}}$. The  $n^{th}$ derivative of $h$ is equal to 
\begin{equation}h^{(n)}(t)=(d/2) (2\rho)^n (1+\rho^2-2 \rho t)^{-n} (n-1)!.\end{equation}
Thus, the Maclaurin series expansions of $g$, for $|t|\leq 1$ is given by
\begin{equation} \label{eq:a1}
\begin{array}{ll}
h(t)& =\log \frac{1-\rho^2}{ (1+\rho^2)^{d/2}}  + \sum_{n=1}^{\infty} h^{(n)}(0) \frac{t^n}{n!} \\
&=\log \frac{1-\rho^2}{ (1+\rho^2)^{d/2}}  + \sum_{n=1}^{\infty} (d/2) (\frac{2\rho}{1+\rho^2})^n ~\frac{t^n}{n}. \\
\end{array}
\end{equation}
The second term in (\ref{eq:a1}) is,
\begin{equation} \label{eq:a2}
\begin{array}{lll}
\sum_{n=1}^{\infty} (d/2) (\frac{2\rho}{1+\rho^2})^n \frac{t^n}{n} &=\frac{d\rho}{1+\rho^2} t+\sum_{n=2}^{\infty} (d/2) (\frac{2\rho}{1+\rho^2})^n \frac{t^n}{n} \\
&\leq \frac{d\rho}{1+\rho^2} t+(d/2)\sum_{n=2}^{\infty} (\frac{2\rho}{1+\rho^2})^n \frac{1}{n} & {\rm since~} |t|\leq1 \\
&=  \frac{d\rho}{1+\rho^2} t+(d/2)\{\sum_{n=1}^{\infty} (\frac{2\rho}{1+\rho^2})^n \frac{1}{n} -\frac{2\rho}{1+\rho^2}\}\\
&=  \frac{d\rho}{1+\rho^2} t+(d/2)\{-\log(1-\frac{2\rho}{1+\rho^2}) -\frac{2\rho}{1+\rho^2}\} &{\rm since~} \log(1-x)=-\sum_{n=1}^{\infty} x^n/n \\
& =  \frac{d\rho}{1+\rho^2} t-(d/2)\log(\frac{1+\rho^2-2\rho}{1+\rho^2}) -\frac{d\rho}{1+\rho^2}. \\
\end{array}
\end{equation}
Let $t={\bf x}.\mbox{\boldmath$\mu$}$, from (\ref{eq:a1}) and (\ref{eq:a2}), 
\begin{equation} \label{eq:a3}
\begin{array}{lll}
\log f({\bf x}|\rho, \mbox{\boldmath$\mu$}) &= h({\bf x}.\mbox{\boldmath$\mu$})-\log \omega_d\\
&\leq \log \frac{1-\rho^2}{ (1+\rho^2)^{d/2}}  + \frac{d\rho}{1+\rho^2} {\bf x}.\mbox{\boldmath$\mu$}-(d/2)\log(\frac{1+\rho^2-2\rho}{1+\rho^2}) -\frac{d\rho}{1+\rho^2} -\log \omega_d\\
&=\log \frac{1+\rho}{ (1- \rho)^{d-1}}  + \frac{d\rho}{1+\rho^2} {\bf x}.\mbox{\boldmath$\mu$}-\frac{d\rho}{1+\rho^2}-\log \omega_d.\\
\end{array}
\end{equation}
Let $\kappa_\rho=\frac{d\rho}{1+\rho^2}$ and 
\begin{equation} \label{eq:a4}
M_\rho=(\frac{1}{c_d(\kappa_\rho) \omega_d \exp(\kappa_\rho )} )
(\frac{1+\rho}{(1-\rho)^{d-1}} ), \end{equation} 
then
\[\log \frac{1+\rho}{ (1- \rho)^{d-1}} -\frac{d\rho}{1+\rho^2}-\log \omega_d=\log c_d(\kappa_\rho)+ \log M_\rho, \]
and so
\begin{equation} \log f({\bf x}|\rho, \mbox{\boldmath$\mu$}) \leq \log M_\rho +\log c_d(\kappa_\rho)+ \kappa_\rho {\bf x}.\mbox{\boldmath$\mu$},\end{equation}
or equivalently,
\begin{equation}
f({\bf x}|\rho, \mbox{\boldmath$\mu$})< M_\rho ~ g(x| \kappa_\rho ,\mbox{\boldmath$\mu$}).\end{equation}\\

{\bf \Large Section B: Calculations Associated with the Estimation of the Number of Clusters} \\

Suppose  ${\hat G}$ is the fitted mixture  model  with density function
\begin{equation}{\hat g}({\bf x})=\sum_{k=1}^M {\hat \pi}_{k} K_{\hat \rho_k}({\bf x}, \mbox{\boldmath$\hat \mu$}_k). \end{equation}
and  $\hat F$ is a nonparametric estimator of the true $F$. Let $F_n(t)=1/n \sum_{i=1}^n I(X_i\leq t)$ be the empirical distribution function of the observations $X_1, \cdots , X_n$ assigning mass $1/n$ to each of the $X_i$'s, then  the kernel density estimator $\hat f$ of the density, is given by
${\hat f({\bf x})} = 1/n \sum_{i=1}^n  K ({\bf x},{\bf x_i}).$
The empirical distance between the fitted mixture model ${\hat G}$ and $\hat F$, based on the Poisson kernel $K_\beta ({\bf x, y})$, is given by
\begin{equation}\label{eq:Vn} D_K(\hat F,\hat G) =1/n^2 \sum_{k=1}^M K_\beta^{ctr(\hat G)} ({\bf x_i},{\bf x_j})=1/n^2 \sum_{k=1}^M {\hat \pi}_{k}  \sum_{i=1}^n  \sum_{j=1}^n  K^{ctr(K_{\hat \rho_k})}({\bf x_i},{\bf x_j}),
\end{equation}
where $K_\beta^{ctr(G)}$ is the $G$-centered kernel defined by
\begin{equation}\label{eq:a2}
K^{ctr(G)} (\textbf{s,t})=K(\textbf{s,t})- K(\textbf{s},G)-K(G,\textbf{t})+K(G,G), \end{equation} 
where $K(x,G)=\int K({\bf x},{\bf y}) dG({\bf y}),$ and $ K(G,G)=\int \int K({\bf x},{\bf y}) dG({\bf x})dG({\bf y})$ (see \cite{Lindsay08}, and \cite{Lindsay14}).\\

\noindent 
We note that, for Poisson kernels $K_\rho ({\bf x},{\bf y})$  and $K_\beta ({\bf y},{\bf z})$  defined on $S^{d-1}\times S^{d-1}$,
\begin{equation}\int_{S^{d-1}}K_\rho ({\bf x},{\bf y})K_\beta ({\bf y},{\bf z}) d\sigma({\bf y}) = K_{\rho \beta}({\bf x},{\bf z}). \end{equation}
and so, 
\begin{equation}\label{eq:a2}
K^{ctr(K_{\hat \rho_k})} (\textbf{s,t})=K_\beta(\textbf{s,t})- 
	K_{\beta \hat\rho_k}(\textbf{s},\mbox{\boldmath$\hat \mu$}_k)-K_{\beta \hat \rho_k}(\mbox{\boldmath$\hat \mu$}_k,\textbf{t})+K_{\beta\hat \rho^2_k}(\mbox{\boldmath$\hat \mu$}_k,\mbox{\boldmath$\hat \mu$}_k), \end{equation} 

Therefore,
\begin{equation}D_K(\hat F,\hat G)=   1/n^2 \sum_{i=1}^n  \sum_{j=1}^n  K_\beta ({\bf x_i},{\bf x_j}) -2/n \sum_{k=1}^M  \sum_{i=1}^n  {\hat \pi}_{k}  K_{\beta \hat \rho_k}({\bf x_i},\mbox{\boldmath$\hat \mu$}_k) + \sum_{k=1}^M   {\hat \pi}_{k}
	K_{\beta \hat \rho^2_k}(\mbox{\boldmath$\hat \mu$}_k,\mbox{\boldmath$\hat \mu$}_k).
\end{equation}

\begin{table}[h!]
	Table A1: Mean empirical distance (ED) with various values of the tuning parameter $\beta$ (ED1, ED2, ED5 corresponding to $\beta=0.1, 0.2 ,0.5$, respectively), $AIC, BIC$ and log-likelihood (loglike) values as a function of the number of clusters $M$. The true number of clusters is three. Data were generated from an equally weighted mixture of Poisson kernel-based densities (PKBD($\rho$)) with mean vectors $(1,0,0),(0,1,0),(0,0,1)$, for various values of $\rho$. The sample size is 100 and the number of Monte Carlo replications is 50. The dimension of the data is three.	{\scriptsize
		\begin{center} 
			\begin{tabular}{|l|l|cccccccc|} \hline
				$\rho$ & Method & $M=2$ & $M=3$ & $M=4$ & $M=5$ & $M=6$ & $M=7$ & $M=8$ & $M=9$ \\	\hline
		0.9	&	ED1	&	0.001060	&	0.000255	&	0.000249	&	0.000249	&	0.000225	&	0.000215	&	0.000218	&	0.000201	\\
		&	ED2	&	0.003682	&	0.000789	&	0.000792	&	0.000787	&	0.000725	&	0.000706	&	0.000714	&	0.000652	\\
		&	ED5	&	0.025608	&	0.003076	&	0.003258	&	0.003432	&	0.003320	&	0.003230	&	0.003334	&	0.002958\\
		&	AIC	&	399.560800	&	218.154000	&	200.368800	&	185.927700	&	173.338000	&	161.372900	&	149.167900	&	137.363600\\
		&	BIC	&	404.771200	&	225.969500	&	210.789400	&	198.953500	&	188.969000	&	179.609100	&	170.009300	&	160.810100\\
		&	loglike	&	-197.780420	&	-106.077000	&	-96.184380	&	-87.963840	&	-80.668990	&	-73.686430	&	-66.583970	&	-59.681800	\\
		0.8	&	ED1	&	0.000821	&	0.000243	&	0.000224	&	0.000210	&	0.000209	&	0.000212	&	0.000194	&	0.000213	\\
		&	ED2	&	0.002759	&	0.000763	&	0.000709	&	0.000665	&	0.000662	&	0.000677	&	0.000630	&	0.000692	\\
		&	ED5	&	0.015587	&	0.002811	&	0.002842	&	0.002820	&	0.002762	&	0.002867	&	0.002735	&	0.002963	\\
		&	AIC	&	467.213500	&	391.504100	&	374.416700	&	360.700100	&	347.661900	&	333.475000	&	320.117600	&	307.077800	\\
		&	BIC	&	472.423900	&	399.319600	&	384.837400	&	373.726000	&	363.293000	&	351.711100	&	340.959000	&	330.524300	\\
		&	loglike	&	-231.606800	&	-192.752100	&	-183.208400	&	-175.350100	&	-167.831000	&	-159.737500	&	-152.058800	&	-144.538900	\\
		0.7	&	ED1	&	0.000467	&	0.000177	&	0.000168	&	0.000159	&	0.000145	&	0.000137	&	0.000145	&	0.000143	\\
		&	ED2	&	0.001554	&	0.000560	&	0.000535	&	0.000516	&	0.000467	&	0.000447	&	0.000474	&	0.000471	\\
		&	ED5	&	0.007911	&	0.002419	&	0.002399	&	0.002417	&	0.002244	&	0.002208	&	0.002349	&	0.002288	\\
		&	AIC	&	504.837400	&	476.553100	&	460.622900	&	445.715900	&	432.538000	&	417.711400	&	403.982200	&	390.086200	\\
		&	BIC	&	510.047800	&	484.368600	&	471.043600	&	458.741700	&	448.169100	&	435.947600	&	424.823600	&	413.532800	\\
		&	loglike	&	-250.418700	&	-235.276600	&	-226.311500	&	-217.857900	&	-210.269000	&	-201.855700	&	-193.991100	&	-186.043100	\\
		0.6	&	ED1	&	0.000308	&	0.000158	&	0.000136	&	0.000142	&	0.000128	&	0.000122	&	0.000128	&	0.000127\\
		&	ED2	&	0.001014	&	0.000509	&	0.000445	&	0.000460	&	0.000423	&	0.000402	&	0.000430	&	0.000423\\
		&	ED5	&	0.004997	&	0.002590	&	0.002306	&	0.002330	&	0.002217	&	0.002172	&	0.002270	&	0.002239	\\
		&	AIC	&	523.083100	&	512.247600	&	497.433900	&	481.920300	&	468.300900	&	453.975300	&	439.565100	&	425.336100	\\
		&	BIC	&	528.293400	&	520.063100	&	507.854600	&	494.946100	&	483.931900	&	472.211500	&	460.406500	&	448.782700\\
		&	loglike	&	-259.541600	&	-253.123800	&	-244.716900	&	-235.960100	&	-228.150500	&	-219.987700	&	-211.782600	&	-203.668100	\\
		0.5	&	ED1	&	0.000206	&	0.000137	&	0.000125	&	0.000127	&	0.000125	&	0.000118	&	0.000116	&	0.000121	\\
		&	ED2	&	0.000704	&	0.000457	&	0.000427	&	0.000424	&	0.000419	&	0.000396	&	0.000389	&	0.000414	\\
		&	ED5	&	0.003993	&	0.002675	&	0.002556	&	0.002433	&	0.002355	&	0.002279	&	0.002232	&	0.002306	\\
		&	AIC	&	535.879400	&	528.179600	&	512.189100	&	498.565500	&	485.277000	&	472.247400	&	457.071100	&	443.147400\\
		&	BIC	&	541.089800	&	535.995100	&	522.609800	&	511.591400	&	500.908000	&	490.483500	&	477.912500	&	466.593900	\\
		&	loglike	&	-265.939700	&	-261.089800	&	-252.094600	&	-244.282800	&	-236.638500	&	-229.123700	&	-220.535600	&	-212.573700	\\
		0.4	&	ED1	&	0.000094	&	0.000083	&	0.000085	&	0.000078	&	0.000078	&	0.000094	&	0.000095	&	0.000093	\\
		&	ED2	&	0.000361	&	0.000311	&	0.000304	&	0.000289	&	0.000285	&	0.000331	&	0.000332	&	0.000325\\
		&	ED5	&	0.002846	&	0.002435	&	0.002196	&	0.002175	&	0.002037	&	0.002146	&	0.002093	&	0.002053	\\
		&	AIC	&	546.491600	&	534.885100	&	522.696700	&	506.722600	&	495.305400	&	478.367700	&	464.698200	&	451.671700	\\
		&	BIC	&	551.701900	&	542.700700	&	533.117400	&	519.748400	&	510.936400	&	496.603900	&	485.539500	&	475.118200	\\
		&	loglike	&	-271.245800	&	-264.442600	&	-257.348400	&	-248.361300	&	-241.652700	&	-232.183800	&	-224.349100	&	-216.835800	\\
		0.3	&	ED1	&	0.000061	&	0.000062	&	0.000062	&	0.000065	&	0.000065	&	0.000076	&	0.000081	&	0.000084\\
		&	ED2	&	0.000260	&	0.000245	&	0.000238	&	0.000245	&	0.000238	&	0.000270	&	0.000291	&	0.000297	\\
		&	ED5	&	0.002533	&	0.002226	&	0.002036	&	0.001986	&	0.001878	&	0.001922	&	0.001971	&	0.001964	\\
		&	AIC	&	551.339200	&	541.158400	&	529.330600	&	516.341700	&	502.805800	&	490.387600	&	476.904000	&	461.691700	\\
		&	BIC	&	556.549500	&	548.973900	&	539.751300	&	529.367600	&	518.436900	&	508.623700	&	497.745300	&	485.138300	\\
		&	loglike	&	-273.669600	&	-267.579200	&	-260.665300	&	-253.170900	&	-245.402900	&	-238.193800	&	-230.452000	&	-221.845900	\\
		0.2	&	ED1	&	0.000044	&	0.000049	&	0.000050	&	0.000057	&	0.000057	&	0.000063	&	0.000068	&	0.000074\\
		&	ED2	&	0.000208	&	0.000213	&	0.000209	&	0.000231	&	0.000227	&	0.000241	&	0.000258	&	0.000277\\
		&	ED5	&	0.002419	&	0.002229	&	0.002090	&	0.002102	&	0.002043	&	0.002009	&	0.002031	&	0.002101	\\
		&	AIC	&	555.764300	&	543.520200	&	529.447500	&	515.932700	&	499.330200	&	487.554000	&	474.111900	&	458.492200	\\
		&	BIC	&	560.974600	&	551.335700	&	539.868200	&	528.958600	&	514.961200	&	505.790200	&	494.953200	&	481.938700	\\
		&	loglike	&	-275.882100	&	-268.760100	&	-260.723700	&	-252.966400	&	-243.665100	&	-236.777000	&	-229.055900	&	-220.246100	\\
		 \hline
			\end{tabular}	
		\end{center}}
	\end{table}
	
\begin{figure}[h!]
	\begin{center}
		\includegraphics[width=0.7\textwidth]{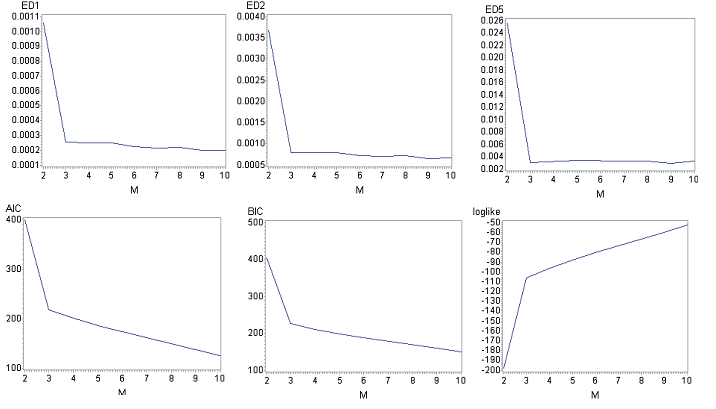}
	\end{center}
	{\footnotesize Figure A1: Empirical Densities Distance Plots (ED) for $\beta=0.1, 0.2 ,0.5$ (ED1, ED2, ED5) and AIC, BIC and log-likelihood plots as a function of the number of clusters $M$.  Data are  generated from a mixture of equally weighted PKBD($\rho=0.9$).  Sample size is 100, the true number of clusters is three, and the data dimension  is three.}
\end{figure}

\begin{figure}[h!]
	\begin{center}
		\includegraphics[width=0.7\textwidth]{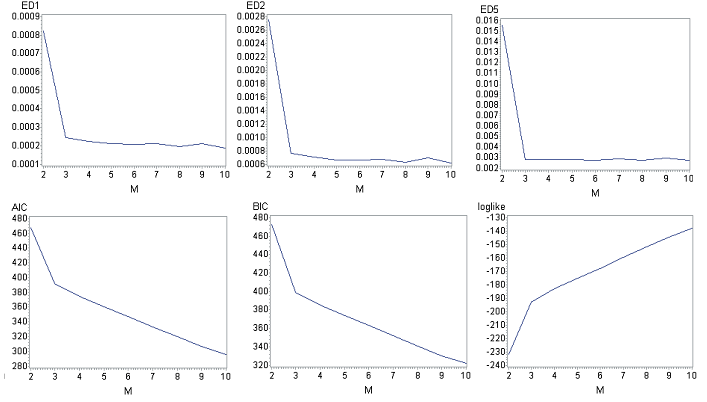}
	\end{center}
	{\footnotesize Figure A2: Empirical Densities Distance Plots (ED) for $\beta=0.1, 0.2 ,0.5$ (ED1, ED2, ED5) and AIC, BIC and log-likelihood plots as a function of the number of clusters $M$.  Data are generated from a mixture of equally weighted PKBD($\rho=0.8$).  Sample size is 100, the true number of clusters is three. Data dimension equals three.}
\end{figure}	
	
\begin{figure}[h!]
		\begin{center}
			\includegraphics[width=0.7\textwidth]{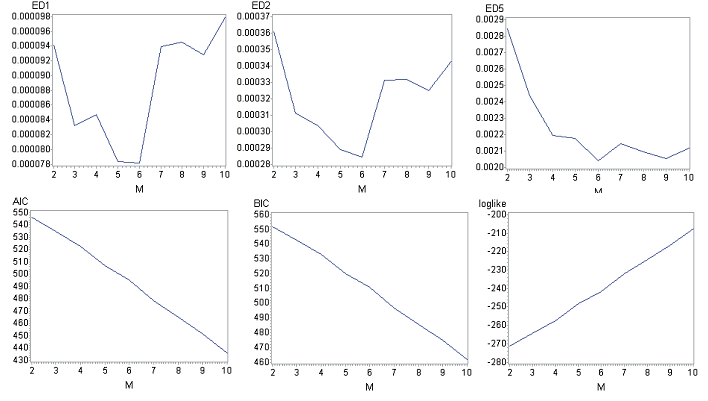}
		\end{center}
		{\footnotesize Figure A3: Empirical Densities Distance Plots (ED) for $\beta=0.1, 0.2 ,0.5$ (ED1, ED2, ED5) and AIC, BIC and log-likelihood plots as a function of the number of clusters $M$.  Data are generated from a mixture of equally weighted PKBD($\rho=0.4$).  Sample size is 100, the true number of clusters is three. Data dimension equals three.}
	\end{figure}
\newpage
Table A1 and Figures A1-A3 show the values of the empirical distance and the empirical densities distance plots for a mixture of three equally weighted PKBD ($\rho$) densities with centers $(1,0,0),(0,1,0), (0,0,1)$, and three values of tuning parameter $\beta$.   The distance plots estimate, in general, correctly the number of true clusters for different values of $\rho$. Specifically, when $\beta=0.1, 0.2 $ the true number of clusters is estimated correctly by the distance plots for all values of $\rho \in [0.2,0.9]$. However, when $\beta=0.5$ the corresponding distance plot for $\rho\in [0.2,0.4]$ appears to be oversmoothed and as a result it overfits the mixture model, selecting a larger number of clusters. More work is needed to understand the impact of the selection of the tuning parameter $\beta$ on the number of identified clusters.\\

{\bf \Large Section C: Additional Simulations and Real Data Examples }\\

\noindent
{\bf \large Evaluation Measures Used in Section 8} \\

Performance of each algorithm is measured by macro-precision, macro-recall \citep{Modha} and also by the adjusted Rand index (ARI; \cite{Hubert}).
Suppose $u_1, \cdots, u_c$ are the true classification classes. For a given clustering,  let $a_t$ denote the number of data objects that are correctly  assigned to the class $u_t$,  $b_t$ denote the data objects that are incorrectly assigned to the class $u_t$, and $c_t$ denote the data objects that are incorrectly rejected from the class  $u_t$. The precision and recall are defined as $p_t=\frac{a_t}{a_t+b_t}$ and $r_t=\frac{a_t}{a_t+c_t}$ for $1 \leq t \leq c$. The macro-precision, and  macro-recall, are the averages across classes of the precisions and recalls. \\

\noindent
{\bf \large Tables for Simulation Study II} \\

\begin{table}[h!]
 Table A2: Macro-precision (M-P), macro-recall (M-R) and Adjusted Rand Index (ARI) as a function of the sample size and dimension. Data are generated from a mixture of a vMF ($\kappa$) and a uniform distribution of equal proportions. Number of Monte Carlo replications is 100. The standard error of the estimates is reported in parenthesis.	{\footnotesize
		\begin{center}		
			\begin{tabular}{|c|c|c|l|l|l l l|}
				\hline 
				N & Dim & $\pi_1$ & Components & Eval.	&  mix-PKBD &  mix-vMF & Spkmeans\\ \hline 	
				&  & & 1 vMF  & M-P & 0.940	(0.02) &	0.984	(0.01) &	0.839	(0.01)
				\\
				200 & 5 & 0.5 & ($\kappa=40$) & M-R & 0.932	(0.02) &	0.984	(0.01) &	0.762	(0.02)	
				\\
				&  &  & 1 uniform  & ARI &0.746	(0.08) &	0.936	(0.04)&	0.273	(0.05) \\
				\hline
				&  & & 1 vMF  & M-P & 0.976	(0.01) &	0.996	(0.01) &	0.844	(0.01) 
				\\
				200 & 15 & 0.5 & ($\kappa=40$) & M-R & 0.974	(0.01) &	0.996	(0.01)	 & 0.772	(0.02)
				\\
				&  &  & 1 uniform  & ARI & 0.899	(0.05) &	0.984	(0.02)&	0.297	(0.05) \\
				\hline
				&  && 1 vMF  & M-P & 0.982	(0.01) &	0.994	(0.01) &	0.848	(0.01)
				\\
				200 & 25 & 0.5 & ($\kappa=40$) & M-R &  0.982	(0.01) &	0.994	(0.01)&	0.781	(0.02)
				\\
				&  &  & 1 uniform  & ARI & 0.928	(0.04) &	0.974	(0.02)&	0.317	(0.04) \\
				\hline
				&  && 1 vMF  & M-P & 0.985	(0.01) &	0.989	(0.01)&	0.854	(0.01)
				\\
				200 & 50 & 0.5 & ($\kappa=40$) & M-R & 0.984	(0.01) &	0.989	(0.01) &	0.794	(0.02)
				\\
				&  &  & 1 uniform  & ARI & 0.937	(0.04) &	0.956	(0.03)&	0.342	(0.05) \\
				\hline
				&  && 1 vMF  & M-P & 0.958	(0.02)&	0.960	(0.01)&	0.852	(0.02)
				\\
				200 & 100 & 0.5 & ($\kappa=40$) & M-R &  0.957	(0.02)&	0.959	(0.01)&	0.804	(0.03)
				\\
				&  &  & 1 uniform  & ARI & 0.834	(0.06) &	0.843	(0.05)&	0.377	(0.07) \\
				\hline  \hline

				&  & & 1 vMF  & M-P & 0.941	(0.007) &	0.985	(0.004) &	0.834	(0.004)			
				\\
				1000 & 5 & 0.5 & ($\kappa=40$) & M-R &	0.932	(0.010) &	0.985	(0.004) &	0.751	(0.010)
				\\
				&  &  & 1 uniform  & ARI & 0.748	(0.033) &	0.939	(0.016) &	0.252	(0.020)
				\\
				\hline
				&  & & 1 vMF  & M-P & 0.976	(0.005) &	0.997	(0.002)&	0.837	(0.005)
				\\
				1000 & 15 & 0.5 & ($\kappa=40$) & M-R & 0.974	(0.006)&	0.997	(0.002)&	0.758(	0.010)
				\\
				&  &  & 1 uniform  & ARI & 0.900	(0.022)&	0.987	(0.008)&	0.266	(0.021)
				\\
				\hline
				&  && 1 vMF  & M-P & 0.985	(0.004)&	0.996	(0.002)&	0.837	(0.004)
				\\
				1000 & 25 & 0.5 & ($\kappa=40$) & M-R &  0.984	(0.004)&	0.996	(0.002)&	0.757	(0.009)
				\\
				&  &  & 1 uniform  & ARI & 0.938	(0.014)&	0.984	(0.006)&	0.265	(0.019)
				\\
				\hline
				&  && 1 vMF  & M-P & 0.986	(0.004)&	0.991	(0.003)&	0.841	(0.005)
				\\
				1000 & 50 & 0.5 & ($\kappa=40$) & M-R & 0.986	(0.004)&	0.991	(0.003)&	0.768	(0.011)
				\\
				&  &  & 1 uniform  & ARI & 0.944	(0.015)&	0.963	(0.013)&	0.285	(0.024)
				\\
				\hline
				&  && 1 vMF  & M-P & 0.966	(0.006)&	0.968	(0.006)&	0.846	(0.005)
				\\
				1000 & 100 & 0.5 & ($\kappa=40$) & M-R &  0.966	(0.006)&	0.968	(0.006)&	0.779	(0.011)
				\\
				&  &  & 1 uniform  & ARI & 0.867	(0.024)&	0.875	(0.022)&	0.310	(0.023)
				\\
				\hline  									\hline 
				\end{tabular}	
			\end{center}}
		\end{table}
The results of Table A2 indicate that when the data are a mixture of vMF and uniform distributions, mix-vMF performs best in terms of ARI, macro-precision \& macro-recall for smaller samples and relatively small dimensions.  For larger samples and higher dimensions mix-PKBD is equivalent to mix-vMF (e.g. when $n=1000$, and dimension is $50$ or $100$).

\begin{table}[h!]
		Table A3: Macro-precision (M-P), macro-recall (M-R) and Adjusted Rand Index (ARI) as a function of the sample size and dimension. Data are generated as a mixture of a PKBD ($\rho$) and uniform distributions of equal proportions. Number of Monte Carlo replications is 25. The standard error of the estimates is reported in parenthesis.		{\footnotesize
		\begin{center}		
			\begin{tabular}{|c|c|c|l|l|l l l|}
				\hline 
				N & Dim & $\pi_1$ & Components & Eval.	&  mix-PKBD &  mix-vMF & Spkmeans\\ \hline 	
				&  && 1 PKBD  & M-P & 0.928 (0.03) & 0.899 (0.03)& 0.824 (0.03)\\
				100 & 5 & 0.5 & ($\rho=0.9$) & M-R & 0.925(0.03) & 0.875 (0.04)& 0.758 (0.04) \\
				&  &  & 1 uniform  & ARI & 0.723 (0.09) & 0.564 (0.11)& 0.267 (0.07)\\
				\hline 
				
				&  && 1 PKBD  & M-P & 0.926	(0.02) & 0.900	(0.03) & 0.821	(0.03)	 \\
				200 & 5 & 0.5 & ($\rho=0.9$) & M-R & 0.924	(0.03) & 0.879 (0.04) &	0.748 (0.04)		 \\
				&  &  & 1 uniform  & ARI & 0.721 (0.07) &	0.576 (0.07) &	0.246 (0.05)
				\\
				\hline 				
				&  & & 1 PKBD  & M-P & 0.928 (0.01) & 0.899 (0.01) & 0.822 (0.01) \\
				1000 & 5 & 0.5 & ($\rho=0.9$) & M-R & 0.927 (0.01) & 0.877 (0.01) & 0.745	(0.01)		 \\
				&  &  & 1 uniform  & ARI & 0.730 (0.03)& 0.568 (0.04) & 0.241 (0.03)\\
				\hline 				
				&  & & 1 PKBD  & M-P & 0.927 (0.004) & 0.897 (0.004) & 0.821 (0.004)
				\\
				3000 & 5 & 0.5 & ($\rho=0.9$) & M-R &  0.926 (0.004)&  0.875 (0.006)&  0.744	(0.006)	
				\\
				&  &  & 1 uniform  & ARI & 0.727 (0.014) & 0.563 (0.019)& 0.239 (0.012) \\
				
				\hline  	\hline			
				&  && 1 PKBD  & M-P & 0.897 (0.02) &	0.894	 (0.02) & 0.826 (0.02) \\
				200 & 25 & 0.5 & ($\rho=0.5$) & M-R & 0.894 (0.02) &	0.885 (0.03) &	0.772 (0.03)\\
				&  &  & 1 uniform  & ARI & 0.622 (0.07) &	0.595 (0.08) &	0.293 (0.06)\\
				\hline		
				&  && 1 PKBD  & M-P & 0.901 (0.009)& 0.894 (0.011) & 0.823 (0.008)\\
				1000 & 25 & 0.5 & ($\rho=0.5$) & M-R &  0.901 (0.009)& 0.886 (0.014) & 0.751 (0.008)\\
				&  &  & 1 uniform  & ARI & 0.642 (0.031) & 0.598 (0.044) & 0.252 (0.022)\\
				\hline \hline			
				&  & & 1 PKBD  & M-P & 0.845 (0.03) &	0.845 (0.05) &	0.760 (0.07)
				\\
				200 & 100 & 0.5 & ($\rho=0.25$) & M-R &	0.837 (0.03) &	0.838 (0.05) &	0.740 (0.06)
				\\
				&  &  & 1 uniform  & ARI & 0.455 (0.09) &	0.457	(0.12) & 0.232 (0.11)
				\\
				\hline  
				&  & & 1 PKBD  & M-P & 0.889 (0.009) & 0.889 (0.009)& 0.818 (0.012)
				\\
				1000 & 100 & 0.5 & ($\rho=0.25$) & M-R & 0.888 (0.009) & 0.888  (0.009)& 0.764 (0.013)
				\\
				&  &  & 1 uniform  & ARI & 0.602 (0.027)& 0.602 (0.027)&  0.276	(0.029)	 \\
				\hline	
			\end{tabular}	
		\end{center}}
	\end{table}
	
	The results of Table A3 indicate that when data are a mixture of PKBD ($\rho$), $\rho=0.9, 0.5, 0.25$ and a uniform, mix-PKBD outperforms mix-vMF and Spkmeans for both small ($N=100$) and larger ($N=3000$) samples and dimension equal to 5. mix-PKBD and mix-vMF have equivalent performance when $\rho=0.5$ and dimension $25$ (for $N=200 ~\&~ 1000$ after considering the standard error given in parenthesis). When the dimension increases to $100$, mix-PKBD and mix-vMF have exactly the same performance.\\
	
\newpage
\noindent
{\bf \large Additional Simulations}

{\it Effect of Number of clusters on the Performance:} Figure A4 plots the different performance measures as a function of the number of clusters. Data are generated from a mixture of $K$ Poisson kernel-based distributions with $\rho=0.8$ on a  $3$-dimensional sphere. 
\begin{figure}[h]
	\begin{center}
		\includegraphics[width=6.5in]{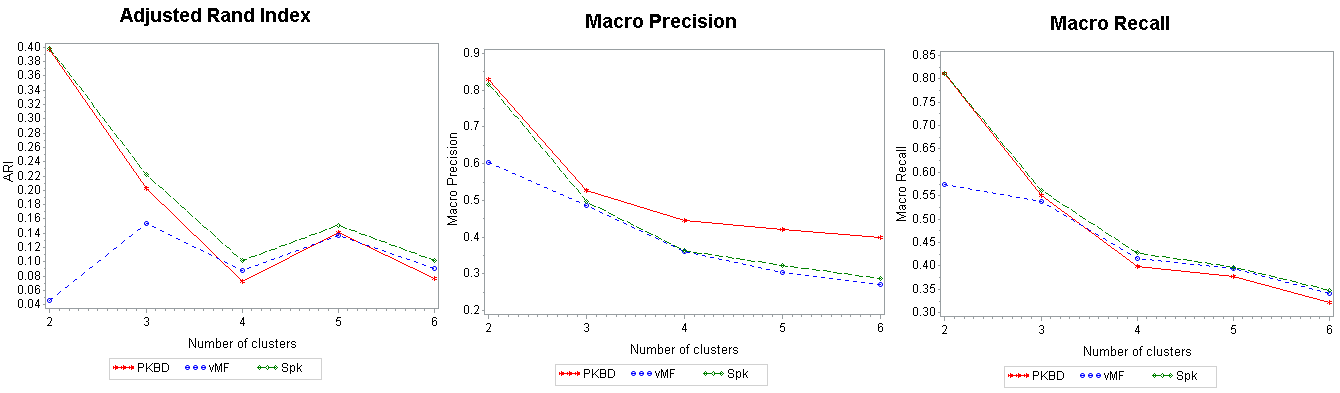}
	\end{center}
 {\footnotesize Figure A4: ARI, Macro-Precision, Macro-Recall of PKBD, vMF and spkmeans algorithms. Data are generated from a mixtures of $k$ equally weighted PKBD ($\rho=0.8$) distribution. Sample size is 200 and the number of Monte Carlo replications is 100. Dimension $d=3$. }
\end{figure}
 
 Figure A5 plots the different performance measures as a function of the number of clusters. Data are generated from a mixture a uniform (50\%) and of $k-1$ equally weighted PKBD with $\rho=0.8$ on a $3$-dimensional sphere. 

\begin{figure}[h]
	\begin{center}
		\includegraphics[width=6.5in]{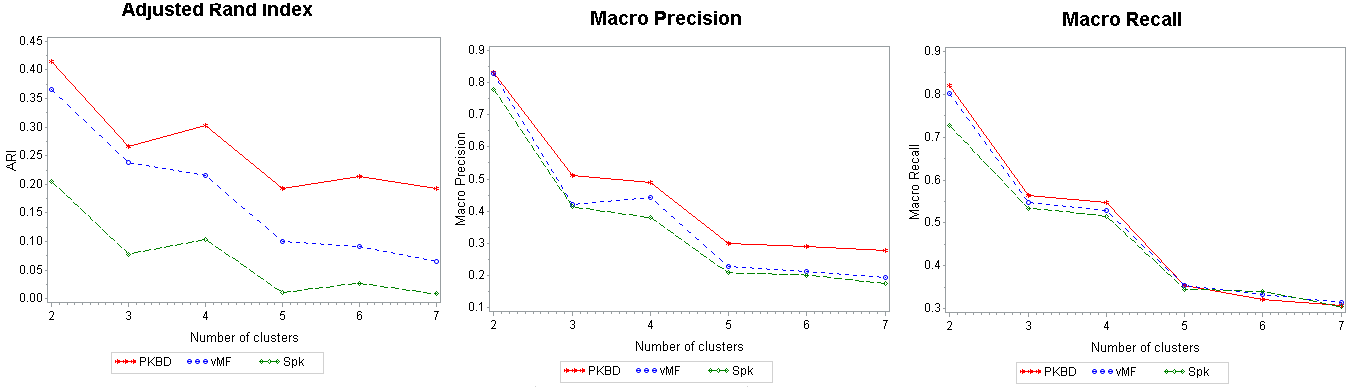}
	\end{center}
	{\footnotesize Figure A5: ARI, Macro-Precision, Macro-Recall of PKBD, vMF and spkmeans algorithms. Data are generated from a mixtures of a uniform (50\%) and $k-1$ equally weighted PKBD ($\rho=0.8$) distribution. Sample size is 200 and the number of Monte Carlo replications is 100. Dimension $d=3$. }
\end{figure}

The plots show that when there is no uniform component in the model the ARI and macro-recall are equivalent for the three models when number of clusters increases but the macro-precision for mix-PKBD is superior than the other two models. However, if one of the mixture components is uniform then mix-PKBD provides for uniformly better results than the other two models  for any number of clusters. \\

{\it Effect of Concentration Parameters of the Components:}
In the previous simulations experiments, we considered a fixed concentration parameter $\rho=0.9$. The results show that the performance of our algorithm is superior in the case when the centers of the PKBD are close and the proportion of the uniform points is high.  In the following simulation experiment, we generate data on a 4 dimensional sphere, from a mixture of 4 distributions, one uniform and three PKBDs. The goal of this experiment is to evaluate the performance of the algorithms for various concentration parameters of the components when the center of the  distributions are fixed. In this experiment, sample size is 200, number of Monte Carlo samples is 100 and percentage of uniform points is 50\%.  We considered three close centers $\mu_1=(0.243, 0, 0.97, 0), \mu_2=(-0.121,  0.21,  0.97,  0),$ and $\mu_3=(-0.121, -0.21, 0.97,  0)$ (corresponding to $a=4$) with the cosine similarity of $0.9118$. Figure A6 shows the performance of the algorithms for different values of the concentration parameters.
The plots show the superior results of mix-PKBD especially for high values of $\rho$.
\begin{figure}[h!]
	\begin{center}
		\includegraphics[width=7in]{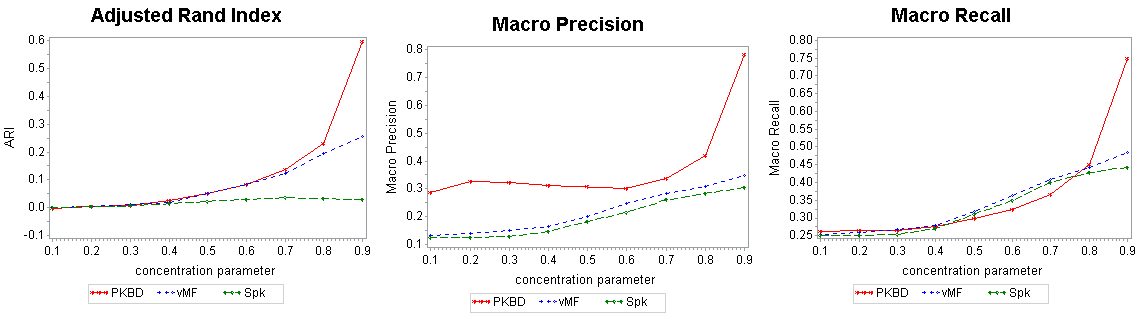}
	\end{center}
{\footnotesize Figure A6: ARI, Macro-Precision, Macro-Recall of PKBD, vMF and Spkmeans algorithms. Data are generated from a mixtures of a uniform (50\%) and $3$ equally weighted PKBD distributions. Sample size is 200 and the number of Monte Carlo replications is 100. Dimension $d=4$. }
\end{figure}

{\it Computational Run Times of the algorithms:}
The goal of the following simulation experiment is to compare the computational run times for the three algorithms. We recorded the run time for each algorithm in seconds to generate mixture of two equally weighted clusters, one uniform and one PKBD, estimate the parameter of interests and gives the class membership. 
Figure A7 plots the run time in seconds of the three algorithms, mix-PKBD, mix-vMF and Spkmeans as a function of the sample size.

\begin{figure}[h!]
	\begin{center}
	\includegraphics[width=0.75\textwidth]{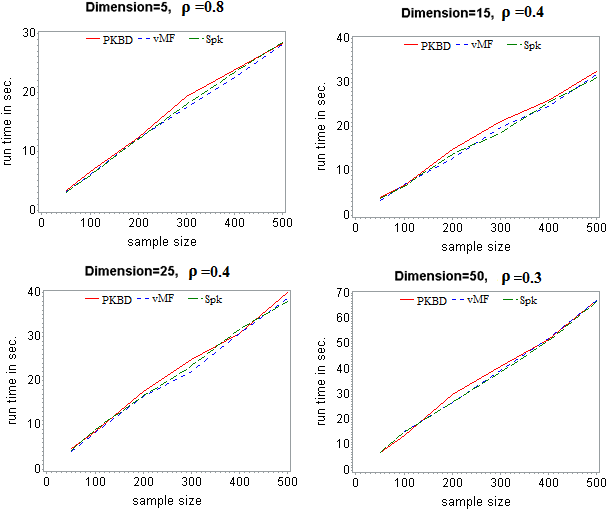}
\end{center}
	{\footnotesize Figure A7: Run time (in seconds) to complete one simulation run for each algorithm. Data are from an equal mixture proportion of a uniform and a PKBD ($\rho$) in different dimensions. The $x$-axis depicts the sample size, the $y$-axis the run time in seconds.}
\end{figure}

The result shows that there is no significant differences in the run time of the algorithms. Similar results are obtained when the mixture densities are uniform and vMF.\\

\newpage
\noindent
{\bf \large Description of the Real Data Sets in Section 8.3}\\

In what follows we briefly describe the real data sets used in this paper. The data sets were selected to exhibit different sample sizes, dimensions, and number of clusters.
\begin{itemize}
	\item [1.]  Text data CNAE-9 was obtained from from UC Machine Learning Repository. This is a data set containing 1080 documents of free text business descriptions of Brazilian companies categorized into a subset of 9 categories cataloged in a table called National Classification of Economic Activities (Classifica\c{c}\~{a}o Nacional de Atividade Econ\^{o}micas - CNAE). The original texts were pre-processed to obtain the current data set. This data set is highly sparse (99.22\% of the matrix is filled with zeros), and contains 1080 documents and 857 attributes. We used Correlated Topic Modeling (CTM) \citep{Blei07, Grun} to reduce the  dimension from 856 words to 50 topics and then used these 50 topics as features.
	
	\item [2.] Text data Congress109 was obtained from "textir" package in R software \citep{Taddy,Taddy16}.  
	This data originally appear in \citet{Gent} and include text of the 2005
	Congressional Record, containing all speeches in that year for members of the United States House
	and Senate. In particular,  Gentzkow and Shapiro record the number of times each of 529 legislators used terms in a  list of 1000 phrases (i.e., each document is a year of transcripts for a single speaker). Associated
	sentiments are rephrased – the two-party vote-share from each speaker's constituency (congressional
	district for representatives; state for senators) obtained by George W. Bush in the 2004 presidential
	election – and the speaker’s first and second common-score values (from http://voteview.com). Full
	parsing and sentiment details are in Taddy (2013; Section 2.1). We used correlated topic modeling \citep{Blei07, Grun} to reduce dimension from 1000 phrases to 100 topics.  The true classes are considered to be the two parties, democrats and republicans. We note that we removed two members of the  independent party from the data set. 
			
	\item [3.] The Crabs data set was obtained from the package "MASS" in R software,  \citep{Camp}. It describes 5 morphological measurements (frontal lobe size, rear width, carapace length, carapace width, body depth) on 50 Leptograpsus crabs each of two color forms (Blue or Orange) and both sexes, of the species Leptograpsus variegatus collected at Fremantle, W. Australia.  We considered the 5 morphological measurements as the features and colors as the true classification.
	
	\item [4.] 		The Quality Assessment of Digital Colposcopies data set was obtained  from UC Machine Learning Repository \citep{Fernan}. 
	The dataset was acquired and annotated by professional physicians at 'Hospital Universitario de Caracas'. 
	The subjective judgments (target variables) were originally done in an ordinal manner (poor, fair, good, excellent) and were discretized in two classes (bad, good). 
	Images were randomly sampled from the original colposcopic sequences (videos). 
	The original images and the manual segmentations are included in the 'images' directory. 
	The dataset has three modalities (i.e. Hinselmann, Green, Schiller). 
	The number of attributes are 69 (62 predictive attributes, 7 target variables). 
	We considered the 62 predictive attributes as the features and the three modalities as the true classification.
	
	\item [5.] The Landsat Multi-Spectral Scanner Image Data (satellite data set) from UC Machine Learning Repository.	
	The database consists of the multi-spectral values of pixels in 3x3 neighborhoods in a satellite image, and the classification associated with the central pixel in each neighborhood. The aim is to predict this classification, given the multi-spectral values. The database is a (tiny) sub-area of a scene, consisting of 82 x 100 pixels. Each line of data corresponds to a 3x3 square neighborhood of pixels completely contained within the 82x100 subareas. Each line contains the pixel values in the four spectral bands (converted to ASCII) of each of the 9 pixels in the 3x3 neighborhood and a number indicating the classification label of the central
	pixel. 
	The data has 6435 rows and 37 columns (x1-x36 continuous variables and class).  The classes are;	red soil,
	cotton crop,	grey soil,	damp grey soil,	soil with vegetation stubble, and	very damp grey soil. 
	
	\item [6.] The household data set was obtained from "HSAUR2" in R software \citep{Ever}. The data is part of a data set collected from a survey on household expenditures and gives the expenses of 20 single men and 20 single women on four commodity groups (housing, food, goods and services).  \citet{Horn} focused only on three of those commodity groups (housing, food and service) to obtain 3-dimensional data for easier visualization. We will focus on all four commodity groups. The scale of measurement of the data is interval.
	\item [7.] Seeds data set was obtained from UC Irvine  Machine Learning Repository. 
	The examined group comprised kernels belonging to three different varieties of wheat: Kama, Rosa and Canadian, 70 elements each, randomly selected for 
	the experiment. High quality visualization of the internal kernel structure was detected using a soft X-ray technique. It is non-destructive and considerably cheaper than other more sophisticated imaging techniques like scanning microscopy or laser technology. The images were recorded on $13\times 18$cm X-ray KODAK plates. Studies were conducted using combine harvested wheat grain originating from experimental fields, explored at the Institute of Agrophysics of the Polish Academy of Sciences in Lublin.
	\item [8.] 	Breast tissue data set, obtained from UC Irvine  Machine Learning Repository, has 9 measurements of 106 samples from 6 different classes of freshly excised tissues.
	Impedance measurements of freshly excised breast tissue were made at the frequencies: 15.625, 31.25, 62.5, 125, 250, 500, 1000 KHz. These measurements plotted in the (real, -imaginary) plane constitute the impedance spectrum from where the breast tissue features are computed. The dataset can be used for predicting the classification of either the original 6 classes or of 4 classes by merging together the fibro-adenoma, mastopathy and glandular classes whose discrimination is not important (they cannot be accurately discriminated anyway). 
	\item [9.] The Birch data set I, contains 3,000 of 2-d data vectors from the three patterns in Figure 3. birch class 1; regular grid, birch class 2; sine curve and birch class 3; random locations \citep{Zhang97}.
	\item [10.] Birch data set II, contains 300,000 data vectors from the same three patterns in Figure A8. 
	\begin{figure}[h!]
		\begin{center}		
		\includegraphics[width=0.4\textwidth]{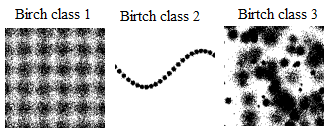}\\
		{\footnotesize Figure A8: The Birch data set, 2-d data vectors from the three patterns}
\end{center}
	\end{figure}
\end{itemize}
The URL page to access these data sets are: (accessed on February 12, 2018)\\
\begin{itemize}
\item [] Text Data CNAE-9: \\ "https://archive.ics.uci.edu/ml/datasets/CNAE-9".\\
\item [] The Congress109 Data Set:\\	"https://cran.r-project.org/web/packages/textir/textir.pdf".\\
\item [] The Crabs Data Set:\\  "https://cran.r-project.org/web/packages/MASS/MASS.pdf". \\
\item [] The Quality Assessment of Digital Colposcopies Data Set:\\ "https://archive.ics.uci.edu/ml/datasets/Quality+Assessment+of+Digital+Colposcopies". \\ 
\item [] The Landsat Multi-Spectral Scanner Image Data Set:\\ "https://archive.ics.uci.edu/ml/datasets/Statlog+(Landsat+Satellite)".\\ 
\item [] The household data set:\\  "https://cran.r-project.org/web/packages/HSAUR2/index.html". \\
\item [] Seeds data set:\\  "https://archive.ics.uci.edu/ml/datasets/seeds". \\
\item [] Breast tissue data set: \\ "http://archive.ics.uci.edu/ml/datasets/breast+tissue".\\
 \item [] The Birtch data set:\\  "https://cs.joensuu.fi/sipu/datasets/".\\
\end{itemize}
\newpage
\begin{table}[h!]	
	Table A4:  Macro-precision, macro-recall and adjusted Rand index for each example
	\begin{center}
		{\footnotesize		\begin{tabular}{|l|l|l|l|c|l l l|}				\hline 
				Data Set & N & K & Dim & Eval.	&  mix-PKBD &  mix-vMF & Spkmeans\\ \hline 
				1. CNAE-9 & 1080 & 9 & reduced  &  M-P &   0.623 & 0.484 & 0.083\\
				& & & to 50	& M-R &  0.493&	0.277 &	0.149\\
				& &	& topics & ARI&   0.215 & 0.050 & 0.001\\
				\hline  
				2. Congress109 &529 & 2&  reduced  &  M-P &   0.684 & 0.601 & 0.269\\
				& & & to 100	& M-R & 	 0.580 &	0.551 &	0.498\\
				& & & topics & ARI&  0.007 &  0.024 & 0.001\\
				\hline  	
				3.	Crabs & 200 & 2 & 5& M-P &  0.949 &	0.532 &	0.531	\\ 
				&  & & &  M-R &  0.950 &	0.530 &	0.584\\
				&&& 	& ARI&  0.809 &	0.001 &	0.001  \\ 
				\hline 
				4. Colposcopies & 287 & 3 & 62 & M-P &   0.919 &	0.713 &	0.657\\	
				&&  &  & M-R &  0.919 & 0.671 &	0.650\\
				& &	 & & ARI& 0.785	 & 0.399 &	0.458\\
				\hline 
				5.	Satellite & 6435 & 6 & 36 &  M-P &   0.699 &	0.580 &	0.610\\
				& &  &	& M-R & 	 0.646 &	0.560 &	0.530\\
				& &	&& ARI& 0.500 &	0.425 &	0.454\\
				\hline 
				6.	Household &	40& 2 & 4& M-P &  0.954&  0.870 &   0.847\\
				& &  &	& M-R &  0.950& 0.825 &    0.825 \\
				& &	&& ARI&0.805 & 0.409 & 0.408\\ 
				\hline  
				7.	Seeds data set &	20 & 3 & 7& M-P &  0.834&  0.768 &   0.698\\
				& &  &	& M-R &  0.814& 0.738 &    0.686 \\
				& &	&& ARI & 0.523 & 0.281 & 0.379\\ 
				\hline 	
				8.	Breast Cancer  & 106 & 6 & 9 & M-P &  0.426 & 	0.436 &	0.403\\ 
				& &  &	& M-R &  0.442 &	0.401 &	0.395\\
				& &	&& ARI& 0.228 &	0.168 &	0.191\\ 
				\hline
				9.	Birch Data set I & 3000 & 3 & 2& M-P & 0.810  & 0.682 &   0.691\\ 
				& &  &	& M-R &   0.812 &	0.690&	 0.704\\
				& &	&& ARI& 0.546 & 0.506  & 0.504 \\ 
				\hline
				10.	Birch Data set II  & 300,000 & 3 & 2& M-P &  0.616  & 0.605 & 	 0.590 \\  
				& &  &	& M-R &  0.619 &	0.642 &	 0.581\\
				& &	&& ARI& 0.307 & 0.264 & 0.250 \\ 
				\hline				
			\end{tabular}}
		\end{center}
	\end{table}
Table A4 presents the performance matrics of the three clustering models, mix-PKBD, mix-vMF \& Spkmeans for each of the aforementioned examples. It is clear from the results that the mix-PKBD model outperforms, in terms of the metrics presented above, mix-vMF and Spkmeans, in most cases (see data set 1, 3, 4, 5, 6, 7 and 9).\\
\newpage

\begin{figure}[h!]
	\begin{center}		
		\includegraphics[width=0.4\textwidth]{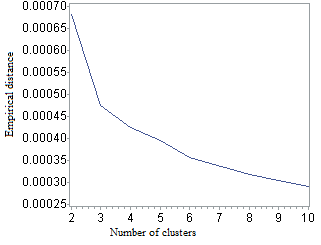}\\
	\end{center}
	{\footnotesize Figure A9: Empirical densities distance plot $(\beta=0.1)$ for the Seeds data. The plot estimates the number of clusters to be three.}
\end{figure}

\begin{figure}[h!]
	\begin{center}		
		\includegraphics[width=0.9\textwidth]{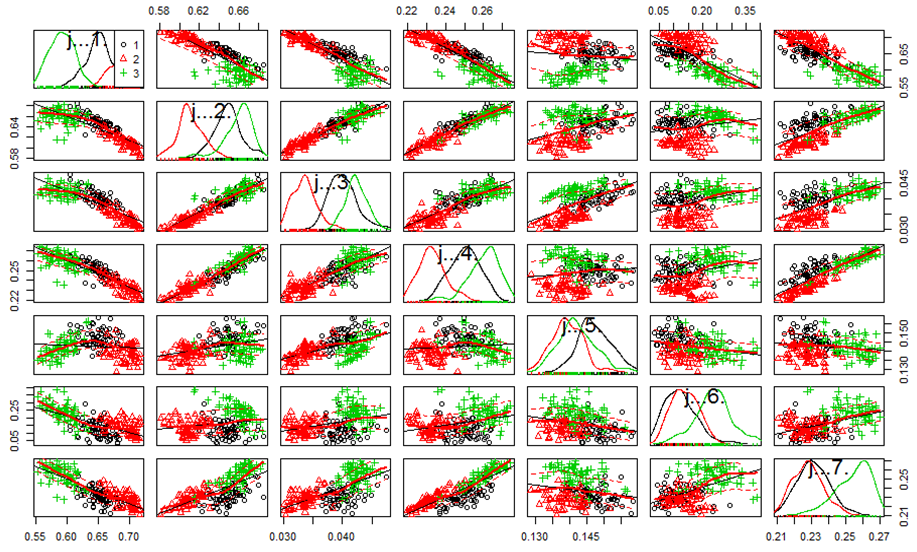}\\
	\end{center}
	{\footnotesize Figure A10: Scatter plot matrix of the Seeds data. The data are of dimension seven. The diagonal contains density estimators of the three clusters in each dimension. The plots indicate a large amount of overlap among these densities.}
\end{figure}
Figures A10 gives the scatter plot matrix for the Seeds data set. 
\newpage
\begin{figure}[h!]
	\begin{center}
		\includegraphics[width=0.64\textwidth]{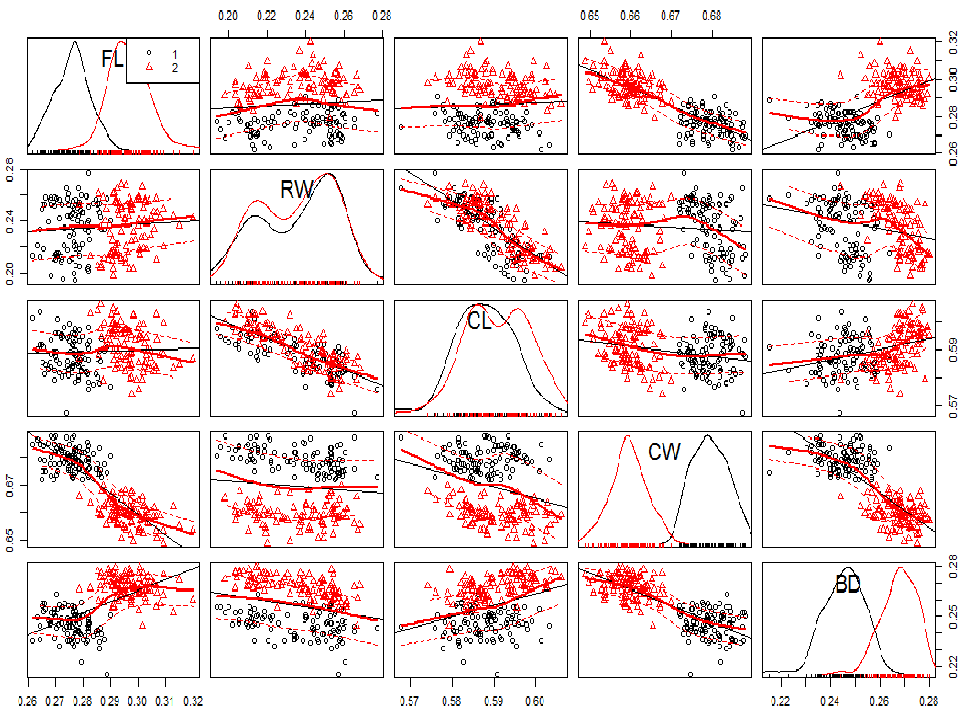}\\
		{\footnotesize 	M-P (PKBD=0.949 vMF=0.532 SpK=0.531)\\
			M-R (PKBD=0.950 vMF=0.530 SpK=0.584) \\ 
			ARI  (PKBD=0.809 vMF=0.000 SpK=0.000) }\\
		Figure A11: Scatter plot matrix for crabs data set by Species
	\end{center}
\end{figure}
\begin{figure}[h!]
	\begin{center}
		\includegraphics[width=0.64\textwidth]{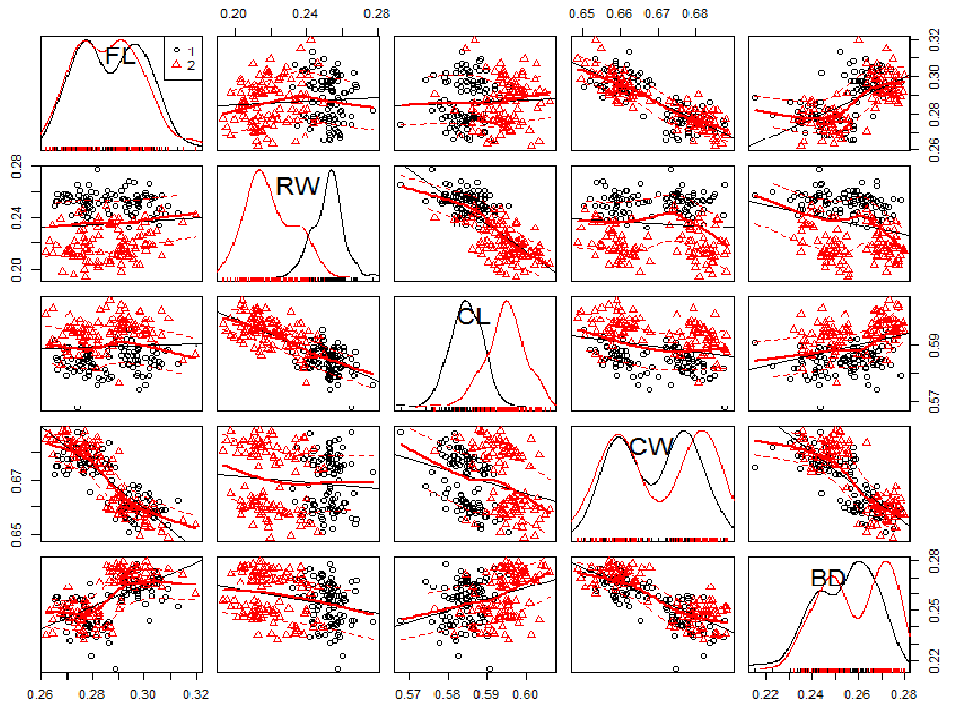}\\
		{\footnotesize 	  M-P (PKBD=0.799 vMF=0.903 SpK=0.908)\\
			M-R (PKBD=0.702 vMF=0.880 SpK=0.888)\\ 
			ARI (PKBD=0.199 vMF=0.575 SpK=0.606)}\\
		Figure A12: Scatter plot matrix for crabs data set by Sexes
	\end{center}
\end{figure}
Figures A11-A12 present the scatter plot matrices associated with the Crabs data set. 
\newpage
\begin{figure}[h!]
	\begin{center}		
		\includegraphics[width=0.45\textwidth]{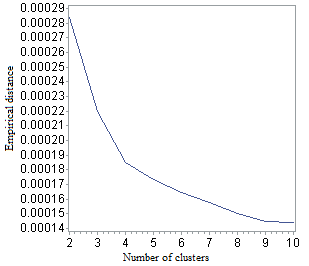}\\
	\end{center}
	{\footnotesize Figure A13: Empirical densities distance plot $(\beta=0.1)$ for the Crabs data set. The plot estimates the number of clusters as four, which  agrees with the estimated number of clusters from the corresponding log-likelihood plot.}
\end{figure}

\begin{figure}[h!]
	\begin{center}		
		\includegraphics[width=0.8\textwidth]{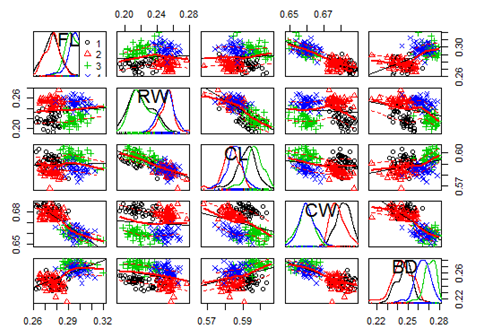}\\
	\end{center}
	{\footnotesize Figure A14: Scatter plot matrix of the Crabs data. The dimension of the data is five.}
\end{figure}
Figure A14 presents the scatter plot matrix associated with the Crabs data set.  
\newpage 
\begin{table}[h!]	
	Table A5:  Performance measures of the different clustering methods when the number of clusters equals four, for the Crabs data set.
	\begin{center}
		{\footnotesize		\begin{tabular}{|l|l l l|}				\hline 
				Clustering Method &   ARI & M-P & M-R\\ \hline
				mix-PKBD &  0.7223 & 0.9042 & 0.8800\\
				mix-vMF & 0.7223 & 0.9042 & 0.8800\\
				Spkmeans & 0.7512 & 0.9130 & 0.8950\\ \hline 
			\end{tabular}}
		\end{center}
	\end{table}
	
{\bf \Large Section D: Additional Tables and Figures }\\

\noindent
{\bf \large Table of the Efficiencies in Section 6} \\

The following table gives the efficiencies of the rejection method for simulating data from Poisson kernel-based distribution with a given concentration parameter $\rho$ using vMF and uniform distribution as upper density, respectively. The efficiencies are defined by $1/M$,  where $M= (\frac{1}{c_d(\kappa_\rho) \omega_d \exp(\kappa_\rho )} )(\frac{1+\rho}{(1-\rho)^{d-1}} ) $ when using vMF distribution and $M= \frac{1+\rho}{(1-\rho)^{d-1}} $ when using uniform distribution.\\

\begin{table}[h]
	Table A6: Efficiencies of the methods using vMF and uniform distributions as upper density 
	\begin{center} 
		\begin{tabular}{|l|l| l | l |}
			\hline 
			Dimension & $\rho$ & Efficiency of vMF & Efficiency of uniform density\\ \hline
			3	&   0.1	&   0.97661 &	0.73636 \\
			3	&	0.4	&	0.60894	&	0.25714	\\
			5	&	0.1	&	0.95492	&	0.59645	\\
			5	&	0.3	&	0.60808	&	0.18469	\\
			10	&	0.1	&	0.90278	&	0.35220	\\
			10	&	0.3	&	0.33698	&	0.03104	\\
			50	&	0.1	&	0.57614	&	0.00521	\\
			50	&	0.2	&	0.08983	&	0.00001	\\
			100	&	0.1	&	0.32863	&	0.00003	\\					\hline	
		\end{tabular}	
	\end{center}
\end{table}

\begin{figure}[h!]
	\begin{center}
	\includegraphics[width=0.9\textwidth]{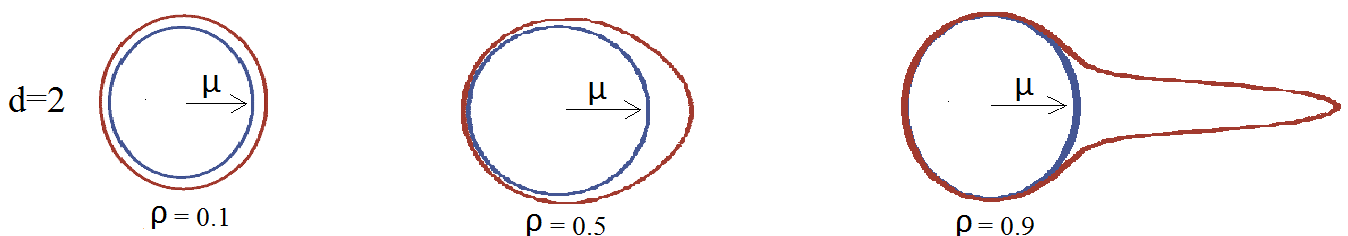}\\
	{\footnotesize	Figure A15: 2-variate Poisson kernel-based distribution for various $\rho$}
	\end{center}
\end{figure}

\begin{figure}[h!]
	\begin{center}
		\includegraphics[width=0.73\textwidth]{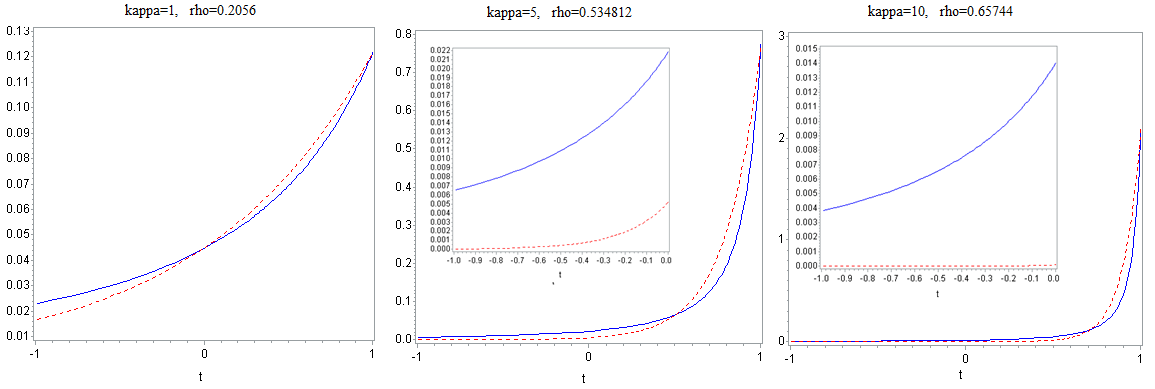}\\
			\end{center}
		{\footnotesize		Figure A16:  Comparison of the PKBD (solid blue) and vMF (dashed red) distributions with the same maximum values, when dimension=$4$ and $t={\bf x}^T\mbox{\boldmath$\mu$}$}
\end{figure}

\begin{figure}[h!]
	\begin{center}
		\includegraphics[width=0.34\textwidth]{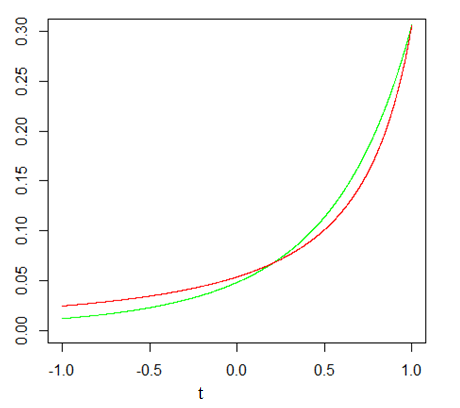}\\
			\end{center}
	{\footnotesize	Figure A17:  Comparison of the PKBD (red) and ESAG (green) distributions with the same maximum values, when dimension=$3$ and $t={\bf x}^T\mbox{\boldmath$\mu$}$, $\delta_1=\delta_2=0$.}

\end{figure}